\documentclass[oldversion]{aa}  

\usepackage{graphicx}
\usepackage{epsfig}
\usepackage{natbib}

\usepackage{txfonts}
\begin{document}
   \title{The extinction law at high redshift and its implications
 \thanks{
 Based on data obtained at the VLT, through the ESO programs
			 074.B-0185
			 and at the Telescopio Nazionale Galileo.}
}
   
   \author{S. Gallerani\inst{1}
\and
	R. Maiolino\inst{1}
          \and
          Y. Juarez\inst{2}
		  \and
		  T. Nagao\inst{3}
\and
A. Marconi\inst{4}
\and
S. Bianchi\inst{5}
		  \and	
R. Schneider\inst{5}
\and
F. Mannucci\inst{5}
\and
T. Oliva\inst{5}
\and
		  C.~J. Willott\inst{6}
\and
  L. Jiang\inst{7}
		  \and
		  X. Fan\inst{7}
}

   \institute{
   INAF-Osservatorio Astronomico di Roma, via di Frascati 33, 00040
              Monte Porzio Catone, Italy
	\and
	Instituto Nacional de Astrofisica, \'Optica y Electr'onica, Puebla, Luis Enrique Erro 1, Tonantzintla, Puebla, 72840, Mexico
	\and
	Research Center for Space and Cosmic Evolution, Ehime University,
	2-5 Bunkyo-cho, Matsuyama 790-8577, Japan
\and
Dipartimento di Fisica e Astronomia, Universit\'a degli Studi di Firenze, Largo E. Fermi 2, Firenze, Italy
\and
	INAF-Osservatorio Astrofisico di Arcetri,
	Largo E. Fermi 5, 50125 Firenze, Italy
	\and
Herzberg Institute of Astrophysics, National Research Council, 5071 West Saanich Rd., Victoria, BC V9E 2E7, Canada
\and
Steward Observatory, 933 N. Cherry Ave, Tucson, AZ 85721-0065, US
             }

   \date{Received ; accepted }
 \abstract{
 We analyze the optical-near infrared spectra of 33 quasars with redshifts $3.9\leq z\leq 6.4$ to investigate the properties of dust extinction at these cosmic epochs. The SMC
 extinction curve has been shown to reproduce the dust reddening of most quasars at $z<2.2$; we investigate whether this curve
 also provides a good description of dust
 extinction at higher redshifts. We fit the observed spectra with
 synthetic absorbed quasar templates obtained by varying the intrinsic slope ($\alpha_{\lambda}$), the absolute
 extinction ($A_{3000}$), and by using a grid of empirical and theoretical extinction curves.
 We find that seven quasars in our sample are affected by substantial extinction ($A_{3000}\geq 0.8$), and characterized by very steep intrinsic slopes ($\alpha_{\lambda}\leq -2.3$). All of the individual quasars
 require extinction curve deviating from that of the SMC, with a tendency to flatten at $\lambda\leq 2000$~\AA~(in the rest frame of the source). However, due to the uncertainties in the individual
 extinction curves the SMC is still (marginally) consistent with the data in most cases.
 We obtain a mean extinction curve at z$>$4, both by performing a simultaneous fit of all quasars and
 by averaging the extinction curves inferred for individual quasars. In the case of broad absorption line quasars (which are generally more absorbed
 by dust and possibly in a younger evolutionary stage), the mean extinction curve deviates from the SMC at a confidence level $\geq 95$\%. The difference between extinction curves in quasars at
 z$>$4 and in quasars at lower redshift is indicative of either a different dust production mechanism
 at high redshift, or a different mechanism for processing dust into the ISM.
 We suggest that the same transitions may also apply to normal, star-forming
 galaxies at z$>$4. In particular, the observed change in the average spectral slope of galaxies
 at z$>$4 may be partially ascribed to a variation in the extinction curve, rather than a lower dust content at
 high redshift.
 In this scenario, the extinction curve inferred at z$>$4 would imply a cosmic star-formation corrected for dust attenuation a factor of $\sim 2$ higher than estimated in the past.
}
\maketitle
\section{Introduction}
Dust represents one of the key ingredients of the Universe by playing a crucial
role in both the formation
and evolution of the stellar populations in galaxies as well as their observability.

Extensive measurements of the
local interstellar extinction exist along hundreds of lines of sight in the Milky Way (MW)
\citep{savagemath}, and along tens of lines of sight in the Large Magellanic Cloud (LMC) and in the
Small Magellanic Cloud (SMC) \citep{fitz}. The wavelength-dependence of the dust extinction and
spectral features in extinction curves are useful for constraining the
size distribution of dust grains and revealing the dust chemical composition. The extinction curves
which characterizing the local Universe have been studied by several authors \citep[e.g., ][]{ccm,mrn}.
In particular, \cite{pei}, by adopting a graphite-silicate grain model, finds that the
MW reddening requires roughly an equal amount of graphite and silicate grains, while the SMC
extinction curve is dominated by silicate grains (see also Weingartner \& Draine 2001), and the LMC
represents an intermediate situation.

Extinction curves that differ from the SMC, LMC, and MW have been obtained in the case of local AGNs.
In the case of mildly obscured AGNs, various authors (Gaskell \& Benker 2007, Maiolino et al. 2001) have found indications of a very flat extinction curve in the UV. Richards
et al. (2003) and Hopkins et al. (2004) found
that mildly reddened QSOs (including broad absorption line, BAL, QSOs) at $z<2.2$ are
characterized by SMC-like
extinction curves. The two results are not inconsistent with each other: while in mildly reddened QSOs the line of
sight passes through a low density and often outflowing medium exposed to the QSO radiation,
in the case of obscured AGNs we are probably piercing through dense clouds where dust growth
may lead to a flattening of the extinction curve 
\citep[similarly to that observed in Galactic clouds, ][]{ccm}.

In the case of star-forming galaxies, it is generally difficult to infer the extinction
curve since, in contrast to quasars, the simple dusty screen case does not apply, a more accurate description being dust mixed with
the emission sources (stars and ionized gas). As a consequence, the observed spectrum heavily depends on the
relative distribution of emitting sources and dust. \cite{calzetti94} inferred an ``attenuation curve''
for star-forming galaxies that determines the effective reddening of the global galaxy spectrum relative
to the case of no extinction, and is flatter than empirical extinction
curves (SMC, LMC, MW). The Calzetti attenuation law can be interpreted in terms of dust mixed with the emitting sources, and with an SMC-like intrinsic extinction curve \citep{inoue05,gordon07}, although the Calzetti law may also be recovered by assuming MW-type dust and different geometries for the emitting sources and dust media \citep{panuzzo07,pierini04}.

At high redshift, the spectral energy distribution (SED) of lyman break galaxies (LBGs) is often modeled with the Calzetti
attenuation curve, although the SMC extinction curve provides a better description for young star-forming
galaxies \citep{reddy10}, indicating that a screen-like dust distribution exists in the latter sources (probably dust associated with the starburst wind). However, \cite{noll09} found that more massive
sources at z$\sim$2 display an absorption dip at 2175~\AA, typical of MW-like extinction curve,
suggesting that these objects are in a more evolved evolutionary stage with dust properties
similar to local disk galaxies.

At the highest redshifts probed so far ($z\sim 6$), the existence of large amounts of dust has been
inferred from mm and submm observations of quasars host galaxies. The far-infrared (FIR) luminosities
of these objects are consistent with thermal emission from warm dust ($T<100~\rm{K}$) with dust
masses $M_{dust}>10^8~\rm{M}_{\odot}$,
assuming that the dominant dust heating mechanism is radiation from young stars 
\citep[e.g. ][]{carilli01,omont01,bertoldi03,beelen06,wang08}.

Many efforts
have been made to explain the enormous amount of dust observed within 1 Gyr of the Big Bang.

Dust formation occurs in environments characterized by temperatures of about $1000-2000$~K and high
density ($n\sim 10^8-10^{15} cm^{-3}$). The proper combination of these conditions is mainly found in
the atmospheres of low and intermediate-mass ($\rm{M}<8~\rm{M}_{\odot}$) evolved stars during the
asymptotic giant branch (AGB) phase, from where dust is transported into the ISM by means of stellar
winds. However, it is generally assumed that the timescales required for the bulk of the AGB population to evolve
(a few $10^8$ to a few $10^9$~yr) are comparable
with the age of the Universe at $z\sim 6$ ($\sim 1$~Gyr); as a consequence, the possibility that evolved
stars can explain the copious amount of interstellar dust observed at these redshifts has been largely
questioned (e.g. Morgan \& Edmunds 2003, but see also Valiante et al. 2009).

Core-collapse type II supernovae (SNeII) may constitute a valid alternative to the dust production at high
redshift. They provide a rapid mechanism of dust enrichment since the lifetimes of their
progenitors are short ($\sim 10^6$ yr), and dust grains are predicted to form in their expanding
ejecta a few hundred days after the explosion. This scenario is corroborated by the infrared emission
and the asymmetric evolution of optical lines observed in the spectra of nearby SNe
\citep{woo93,elm03,sug06,meikle07,kotak06,wesson10},
as well as by infrared and submm observations of SN remnants
\citep{morgan03,rho08,rho09,gomez09}. 
These observations suggest that the total mass of dust produced by SNe is of
the order of a few times $10^{-2}~\rm{M}_{\odot}$ per SN, although these
estimates are disputed. For example, dust yields as high as $\rm 0.1-1.0~\rm{M_{\odot}}/SN$
have been claimed for the SN remnants Cassiopeia A and Kepler
\citep{morgan03,gomez09,dunne09}.

From a theoretical point of view, several authors have proposed
that the ejecta of SNeII (and also pair instability SNe, PISNe)
are an efficient dust formation mechanism at high redshifts
\citep{tf01,sch04a,hira08}. These models predict both the
SNeII/PISNe extinction curves and the masses of dust produced by each class of SN. For
instance, 
SNeII from progenitors with $12-40~\rm{M}_{\odot}$
are predicted to produce dust masses of about $0.1-2~\rm{M}_{\odot}/SN$. The
discrepancy between the mass of dust inferred from most of the observations ($<5\times
10^{-2}~\rm{M}_{\odot}$) and theoretical predictions ($>0.1~\rm{M}_{\odot}$) can be reconciled by
including the effects of the reverse shock, which can destroy a substantial ($\gtrsim$ 80\%)
fraction of the dust formed in the SN ejecta
\citep{bs07,nozawa07,hira08}.

The scenario of SN-like dust formation in the early Universe has been investigated observationally
by studying the extinction curve in z$\sim$6 sources, and in particular by analyzing the reddened quasar SDSSJ1048+46 at $z=6.2$ \citep{m04}, the GRB050904 afterglow at $z=6.3$ \citep{stratta07}
and the GRB071025 at $z\sim5$ \citep{perley}.
These studies indicate that the inferred dust extinction curves differ from those observed at low redshift,
while being in very good agreement with the \cite{tf01} predictions of dust formed in SNe
ejecta. However, it must be mentioned that \cite{lili08} also analyzed the extinction of the GRB050904
host galaxy, finding a somewhat different result from that of \cite{stratta07},
but by assuming for the GRB intrinsic optical continuum a much shallower power-law slope $\beta _{opt}$ 
than that inferred from the X-ray data
(more specifically $\beta _{opt}\sim 0.25~\rm{vs}~\beta _X\sim 1.2$). \cite{zafar10} revised the photometric measurements of GRB050904 reported in the literature and investigated whether this
GRB is significantly absorbed by dust.
   
The question of whether the SN-dust formation is efficient enough to account for
$M_{dust}>10^8~\rm{M}_{\odot}$ at $z\sim 6$ remains a topic of debate. \cite{dwek07} applied the
results of an analytical model for the evolution of the gas, dust, and metals in high-$z$ galaxies to
one of the most distant quasars (SDSSJ1148) at $z=6.4$. Based on the assumption that SNe are the only
available dust suppliers at high redshifts, these authors find that their model can explain the
observed dust mass, provided that the IMF is top-heavy and the dust condensation efficiency is as
high as 100\%; they also suggest that substantial grain growth may occur in the ISM  \citep[see also][]{michal10a,michal10b}.
\cite{val09} are able to reproduce the dust mass observed in SDSSJ1148 at $z=6.4$, under
much less extreme conditions, by means of a chemical evolutionary model that includes the role of
AGB stars as dust suppliers at high redshifts. By comparing their model predictions with
the mass of dust inferred from SDSSJ1148, they find that AGB stars can
account for 50-80\% of the total mass of dust, while SNe may account for the remaining fraction
(50-20\%), depending on the star-formation history of the host galaxy.

Finally, \cite{elvis02} present an alternative to the normally assumed
stellar origin of cosmic dust, by proposing a dust condensation scenario connected to quasar
outflow winds. According to their scenario, the conditions for dust
formation may occur within the clouds responsible for the broad emission lines in expansion
through the quasar outflowing wind (provided that metals, and in particular refractory
elements, are already available). Whether this mechanism can explain the huge amount of
dust at high redshift remains unclear and depends on the quasar outflow rate across its
evolutionary history, the quasar nuclear
metallicity evolution, and the grain condensation efficiency within the outflowing clouds.

To summarize, the dust production mechanism at high-$z$ is still a hotly debated issue. In this work, we
study the properties of dust at $3.9<z<6.4$ by analyzing the optical/NIR spectra of
a sample of $\sim 30$ QSOs. In particular, we investigate whether the dust extinction
inferred from high redshift quasar spectra can be described in terms of the SMC extinction curve,
as for quasars at $z<2.2$ (Richards et al. 2003; Hopkins et al. 2004), or it
reflects variations in the dust grain size distribution and/or in the dust chemical composition. The
outline of this paper is as follows. Section 2 presents the observational data. In Sect. 3, we
describe the properties of the extinction curves adopted in this study. Moreover, we discuss the results of the analysis
of the quasar spectra that require substantial dust reddening and we quantify the deviation of the mean observed extinction curve from the SMC.
The implications of our findings are discussed in Sect. 4, while, in Sect. 5, we summarize the main
results of our study.

\section{Spectroscopic data and sample}

\begin{table*}
\centering
\caption{Sample of high-z quasars}
\label{tab1}
\begin{tabular}{l l l l l}
\hline
 Name             &  z & Ref. & Instrument & Disperser$^{\mathrm{a}}$  \\
\hline
SDSS J000239.4+255035 &  5.80&$^{\mathrm{[1]}}$&NICS-TNG&Amici+IJ  \\
SDSS J000552.3-000656 &  5.85 &$^{\mathrm{[1]}}$&NICS-TNG&Amici\\
SDSS J001714.66-100055.4$^{\mathrm{b}}$ &  5.01&$^{\mathrm{[1]}}$&FORS2-VLT&GRIS150I+27\\
SDSS J012004.82+141108.2$^{\mathrm{b}}$ &  4.73&$^{\mathrm{[1]}}$&NICS-TNG&Amici  \\
SDSS J015642.11+141944.3$^{\mathrm{b}}$ &  4.32&$^{\mathrm{[1]}}$&NICS-TNG&Amici  \\
ULAS J020332.38+001229.2$^{\mathrm{b}}$ & 5.72 & $^{\mathrm{[4]}}$&GST&GNIRS\\
SDSS J023137.6-072855 &  5.41&$^{\mathrm{[1]}}$&FORS2-VLT&GRIS150I+27   \\
SDSS J023923.47-081005.1$^{\mathrm{b}}$ &  4.02&$^{\mathrm{[1]}}$ &NICS-TNG&Amici  \\
SDSS J033829.3+002156 &  5.00&$^{\mathrm{[1]}}$ &FORS2-VLT&GRIS150I+27   \\
SDSS J075618.1+410408 &  5.07&$^{\mathrm{[1]}}$  &NICS-TNG&Amici  \\
APM  08279+5255$^{\mathrm{b}}$ & 3.90 &$^{\mathrm{[1]}}$&NICS-TNG&Amici\\
SDSS J083643.8+005453 &  5.80&$^{\mathrm{[2]}}$ &GST&GNIRS \\
SDSS J085210.89+535948.9$^{\mathrm{b}}$ &  4.22&$^{\mathrm{[1]}}$  &NICS-TNG&Amici  \\
SDSS J095707.67+061059.5 &  5.16 &$^{\mathrm{[1]}}$ &FORS2-VLT&GRIS150I+27  \\
SDSSp J102119.16-030937.2 &  4.70 &$^{\mathrm{[1]}}$  &FORS2-VLT&GRIS150I+27 \\
SDSS J103027.1+052455 &  6.28 &$^{\mathrm{[2]}}$&GST&GNIRS  \\
SDSS J104433.04-012502.2$^{\mathrm{b}}$ & 5.78 &$^{\mathrm{[2]}}$ &GST&GNIRS \\
SDSS J104845.05+463718.3$^{\mathrm{b}}$ & 6.20 &$^{\mathrm{[1]}}$ &NICS-TNG&Amici+IJ +HK\\
SDSS J114816.6+525150 &  6.40  &$^{\mathrm{[1]}}$  &NICS-TNG&Amici\\
SDSS J120441.7-002150 &  5.05  &$^{\mathrm{[1]}}$ &FORS2-VLT&GRIS150I+27 \\
SDSSp J120823.8+001028 &  5.27  &$^{\mathrm{[1]}}$ &FORS2-VLT&GRIS150I+27 \\
SDSS J130608.2+035626 &  5.99   &$^{\mathrm{[2]}}$&GST&GNIRS\\
SDSS J141111.3+121737 &  5.93   &$^{\mathrm{[2]}}$&GST&GNIRS\\
CFHQS J150941.78-174926.8$^{\mathrm{b}}$ & 6.12 &$^{\mathrm{[3]}}$&GST&GNIRS\\
SDSS J160254.2+422823 &  6.07   &$^{\mathrm{[1]}}$  &NICS-TNG&Amici+IJ\\
SDSS J160320.89+072104.5 &  4.39  &$^{\mathrm{[1]}}$  &NICS-TNG&Amici\\
SDSS J160501.21-011220.6$^{\mathrm{b}}$ & 4.92   &$^{\mathrm{[1]}}$ &  NICS-TNG&Amici+IJ\\
SDSS J161425.13+464028.9 &  5.31  &$^{\mathrm{[1]}}$  &NICS-TNG&Amici\\
SDSS J162331.8+311201 &  6.26   &$^{\mathrm{[1]}}$  &NICS-TNG&Amici\\
SDSS J162626.50+275132.4 &  5.20  &$^{\mathrm{[1]}}$  &NICS-TNG&Amici\\
SDSS J163033.9+401210 &  6.06  &$^{\mathrm{[1]}}$  &NICS-TNG&Amici\\
SDSS J220008.7+001744 &  4.77  &$^{\mathrm{[1]}}$  &NICS-TNG&Amici\\
SDSS J221644.0+001348 & 4.99    &$^{\mathrm{[1]}}$  &NICS-TNG&Amici\\
\hline
\end{tabular}

$^{\mathrm{a}}$Resolution: Amici ($R\sim50-75$), IJ ($R\sim300$), GRIS ($R\sim300$), GNIRS ($R\sim800$), HK ($R\sim300$); $^{\mathrm{b}}$ BAL quasars; $^{\mathrm{[1]}}$ This work and Juarez et al. (2009);
$^{\mathrm{[2]}}$ Jiang et al. (2007); $^{\mathrm{[3]}}$ Willott et al. (2007); $^{\mathrm{[4]}}$ Mortlock et al. (2009)
\end{table*}

Most of the spectra used in this study were obtained in various observing runs at the Telescopio
Nazionale Galileo and at the ESO Very Large Telescope. Many of these spectra were already 
partly presented in previous works (Maiolino et al. 2003, 2004b, Juarez et al. 2009). The data
reduction process is presented in these previous papers, and here we summarize only the instrumental setup.
The majority of the spectra were obtained with the TNG near-IR spectrometer
NICS \citep{baffa01} by using the AMICI prism, which allows us to observe the entire 0.9--2.3~$\mu$m spectral
range in a single observation, though at low spectral resolution (R$\sim$50-75, which is enough for
the purposes of investigating the dust extinction through the spectral shape). A few quasars were also
observed with NICS at intermediate dispersion (R$\sim$300) through the IJ and HK grisms (mostly to more clearly resolve a few absorption and emission line features).

We also include a number of quasars at z$<$5
observed with FORS2 by using the  grism GRIS150I, which covers the 0.7-1.0$~\mu$m range for
R$\sim$300. However, in most of these cases the limited rest-frame spectral coverage
does not allow us to constrain the dust reddening.

The selection criterion for quasars at z$>$5 was simply to include all quasars known at the epoch
of the observations and visible during the observing runs. In the redshift range 3.9$<$z$<$5, we gave
preference to known BAL quasars, since these tend to be more often affected by dust extinction, hence more suitable
for investigating the extinction curve. As a consequence,
the large number of BAL quasars in our sample of quasars at
3.9$<$z$<$6.4 simply reflects our selection criterion.

We also include in our sample the near-IR spectra of five z$\sim$6 quasars by Jiang et al. (2007),
which were obtained with a spectral resolution $R\sim 800$ at GEMINI South by using GNIRS,
which provides a simultaneous wavelength coverage from 0.9 to 2.5 $\mu m$. Finally,
we also include the NIR spectrum of 
CFHQS J1509-1749 ($z=6.12$) obtained by Willott et al. (2007) at GEMINI South by using GNIRS
and the NIR spectrum of
ULAS J0203+0012 ($z=5.72$) by Mortlock et al. (2009) at GEMINI North with GNIRS.

All the quasars adopted in our analysis are reported in Table 1 along with important information about the spectroscopic data. The spectra of most quasars are shown in electronic form (Appendix A), except for the
reddened targets (the most interesting cases), which are shown in the next section.

Hereafter, we refer to each object by using only the acronym of the survey and the
first four digits of its right ascension and declination.

\section{Determining the extinction curve at high redshift}

\subsection{The method}\label{method}

The spectrum of each quasar is fitted with the following expression
\begin{equation}
F_{\lambda}=C~F_{\lambda}^{t}~\left ( \frac{\lambda}{3000\AA}\right )^{(1.62+\alpha_{\lambda})}~10^{-\frac{A_{3000}}{2.5}
\frac{A_{\lambda}}{A_{3000}}},
\label{bfeq}
\end{equation}
where C is a normalization constant, $F_{\lambda}^{t}$ is a quasar template spectrum, $\alpha_{\lambda}$ the intrinsic slope
of the unreddened spectrum, $A_{3000}$ the absolute extinction at 3000~\AA, and $A_{\lambda}/A_{3000}$ is the extinction curve
normalized at 3000~\AA, which is derived from the library of empirical and theoretical extinction curves
discussed in the next section. In this study, we adopt the template compiled by Reichard et al. (2003), obtained by
considering 892 quasars classified as non-BAL, whose continuum is described by a power-law with index $\alpha_{\lambda,t}=-1.62$
($F_{\lambda}^{t}\propto \lambda^{\alpha_{\lambda,t}}$).
Therefore, the term
$(\lambda/3000)^{(1.62+\alpha_{\lambda})}$ allows us to force the slope of the template to the value $\alpha_{\lambda}$. We
allow $\alpha_{\lambda}$ to vary in the interval [-2.9; -0.2],
which is the range encompassed by more than 95\% of the QSOs in the SDSS sample of Reichard et al. (2003).
The normalization of the intrinsic quasar spectrum is allowed to vary, with the only constraint being that the intrinsic quasar
bolometric luminosity should not exceed the Eddington value, within a factor of 1.5, to allow for uncertainties in the
bolometric correction \citep{risaliti04}. For each quasar, the Eddington luminosity is inferred from black hole masses
obtained from the literature (Willott et al. 2003; Jiang et al. 2006; Kurk et al. 2007), or estimated from our own spectra
by using the virial calibrations given in \cite{vest06}.

In the fitting procedure, we avoid spectral regions affected by the following emission lines:
Ly$\alpha$+NV [1215.67; 1280]~\AA; OI+SiII
[1292; 1320]~\AA; SiIV+OIV] [1380; 1418]~\AA; CIV [1515; 1580]~\AA; AlIII+CIII] [1845; 1945]~\AA. We also
exclude the region 2210--3000~\AA, since it is generally characterized by a prominent FeII bump. Moreover, in the case of BAL
QSOs, we identify, and discard from the fitting, the associated absorption troughs. Spectral regions with
some instrumental artifacts were also removed.
Finally, we avoid the following regions in the observed frame, which are affected by poor atmospheric transmission: 1.33--1.5 $\mu m$; 1.8--1.95 $\mu m$; $>2.3$ $\mu m$. After selecting the spectral regions to be included in the analysis, we rebin the observed spectra to a
resolution $R\sim 50$, which is about the spectral resolution delivered by the Amici prism in most
NICS-TNG observations. However, we checked that the results do not depend on the binning adopted.

For each quasar observed in our sample, we find the best fitting $\alpha_{\lambda}$, $A_{3000}$, and the extinction curve $A_{\lambda}$ that minimize the $\chi^2$.
Errors in our spectra (especially in the near-IR) are dominated by calibration uncertainties. In particular, the intrinsic
shape of the telluric calibration standard star (used also to calibrate the relative instrumental response) is generally
known with an uncertainty of about $err_{rel}=5\%$, which is therefore assumed as the average uncertainty in our
spectra. In some cases, some rebinned spectra regions are affected by a noise higher than 5\%; in these cases, we rebinned the spectra again to reach noise levels below the calibration uncertainty of 5\%.

Besides the best fit
$\chi^2$ ($\chi^2_{BF}$), we compute the $\chi^2$ for the case where no extinction is assumed (i.e.
$A_{3000}=0$), defined as $\chi^2_{NOext}$.
We claim that a given quasar spectrum requires extinction
if the probability of getting a $\chi^2$ lower than $\chi^2_{Noext}$
is $\gtrsim$ 90\% (i.e. $P(\chi^2_{Noext})\gtrsim 90\%$), and, at the same time,
if the probability of getting a $\chi^2$ lower than $\chi^2_{BF}$ is $\lesssim$ 68\% (i.e. $P(\chi^2_{BF})\lesssim 68\%$).

It should be noted that the method to infer the extinction curve adopted in this paper, i.e. by fitting template spectra
and extinction curves to the observed spectra, differs from the method employed in Maiolino et al. (2004a),
who inferred directly the extinction curve by inverting Eq.~1. Although the latter method has the advantage of not
assuming any shape for the extinction curve {\it a priori}, it does require assumptions about the intrinsic spectrum
of the QSO, in terms of both normalization and slope; in particular, in Maiolino et al. (2004a)
a quasar luminosity close to the Eddington limit was assumed and spectral slopes varying within an interval narrower
than adopted here. In contrast, with the ``fitting method'' used in this paper, the intrinsic spectrum of the
QSO, and in particular its normalization and slope, are quantities that are fitted by the procedure.

\begin{figure*}
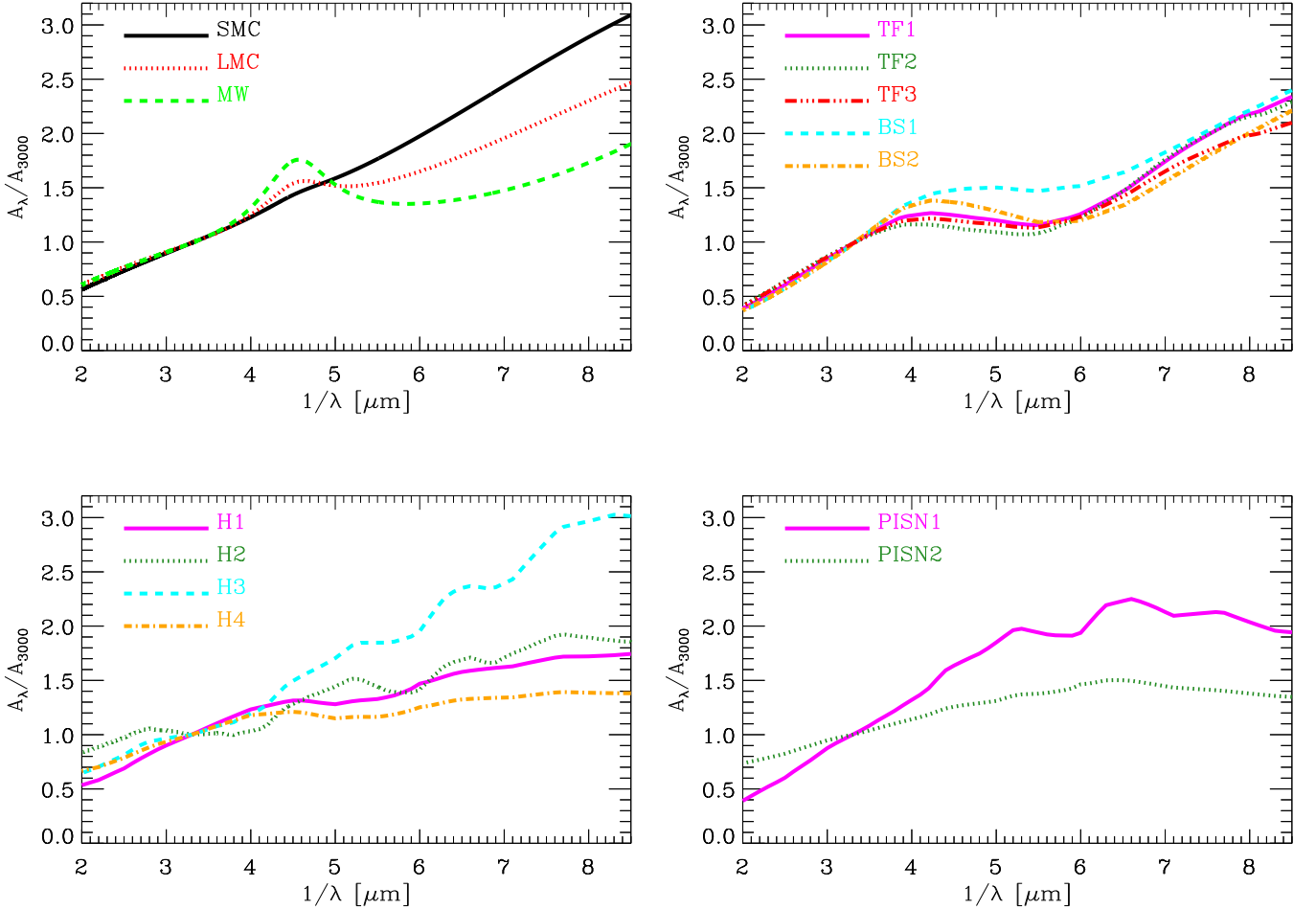

\centerline{
\psfig{figure=14721fg1.ps,width=9.5cm,angle=0}
$\!\!\!\!\!\!\!\!\!\!\!$
\psfig{figure=14721fg2.ps,width=9.5cm,angle=0}
}
\centerline{
\psfig{figure=14721fg3.ps,width=9.5cm,angle=0}
$\!\!\!\!\!\!\!\!\!\!\!$
\psfig{figure=14721fg4.ps,width=9.5cm,angle=0}
}
\caption{{\it Upper left panel}: Empirical curves for dust extinction in the local Universe \citep{savagemath,fitz}. {\it Upper right panel}: Theoretical extinction curves predicted by Type II SNe dust models proposed by Todini \& Ferrara (2001) and Bianchi \& Schneider (2007). {\it Lower left panel}: Results of the Type II SNe dust model by Hirasita et al. (2008). {\it Lower right panel}: Predictions for extinction curves produced by PISNe dust (Hirasita et al. 2008). The main properties of the theoretical models shown in this figure are mentioned in Sect. \ref{secec}.}
\label{ec}
\end{figure*}

\subsection{Library of extinction curves}\label{secec}

To investigate the evolution of the dust properties across cosmic time, we consider a grid of
extinction curves to characterize the extinction produced by dust in the rest-frame wavelength range
$\rm 1000~\AA \leq \lambda \leq 5000~\AA $.
We first consider the empirical curves that describe the dust extinction in the local Universe \citep{savagemath,fitz}, i.e., the SMC, LMC, and MW extinction
curves (Fig.~\ref{ec}, upper left hand panel). 
We then use the extinction curves expected by Type II SNe dust models as
predicted by \cite{tf01}, \cite{bs07}, and
\cite{hira08}. The extinction curves of \cite{tf01}, shown in the right hand upper panel of Fig.
\ref{ec} depend on the metallicity of the SNe progenitors; the right-hand upper panel of Fig.~\ref{ec}
shows the cases of $Z=0$ (TF1, magenta solid line), $Z=Z_{\odot}$ (TF2, green dotted line), and
$Z=10^{-2}Z_{\odot}$ (TF3, red dotted-long-dashed line).
In their model, \cite{bs07} take
into account the possibility that the freshly formed dust in the SNe envelopes can be destroyed and/or
reprocessed by the passage of the reverse shock. They therefore follow the evolution of the dust
condensed from the the time of formation in the ejecta to its survival in the SNe remnants a few thousand years later. The right-hand upper panel of Fig.~\ref{ec} shows the extinction curve predicted by the model
before (BS1, cyan dotted line) and after (BS2, orange dotted-short-dashed line) the reverse shock,
assuming solar metallicity for the SNe progenitor. Both the extinction curves by \cite{tf01}
and \cite{bs07} are averaged over the Salpeter initial mass function.\\ 
In the bottom left-hand panel of Fig. \ref{ec}, we show the results of the models by \cite{hira08} for
dust produced by SNe of progenitor mass 20~$M_{\odot}$. In this work, the effect of the reverse
shock is also considered, for various densities of the ambient medium. Moreover, \cite{hira08} study
two extreme cases for the mixing of the elements constituting the dust grains: the first is the unmixed
case, where the original onion-like structure of elements is preserved in the helium core;
the second is the mixed case, characterized by a uniform mixing of the elements. The solid magenta (dashed cyan)
line shows the mixed case, after the reverse shock, for a medium characterized by hydrogen number
density of $n_H=1~\rm{cm}^{-3}$ ($n_H=0.1~\rm{cm}^{-3}$). Hereafter, we label
these models as H1 and H3, respectively. The
dotted green line represents the unmixed case for freshly formed dust grains (H2), while the
dotted-dashed orange line shows the predictions for the unmixed case after the reverse shock, for
$n_H=0.1~\rm{cm}^{-3}$ (H4).\\ 
\cite{hira08} also investigated the case in which the extinction
is caused by dust produced by pair instability supernovae (PISNe) with a progenitor mass of
170~$M_{\odot}$. In the right-hand bottom panel of Fig. \ref{ec}, we show the results of their analysis for a
medium characterized by $n_H=1~\rm{cm}^{-3}$ in the mixed (PISN1, solid magenta line) and unmixed (PISN2,
dotted green line) cases.\\
We note that our analysis does not really aim to distinguish between different SN-dust progenitors using the inferred extinction curve. We more simply aim to adopt a library of extinction curves that are physically
plausible.

Predictions for the extinction curve produced by dust formed in quasar winds (Elvis et al. 2002)
have not yet been developed, hence this scenario cannot be included in our analysis.
\begin{table*}
\begin{center}
\caption{Best-fit model parameters for reddened quasars}
\label{tab2s}
\begin{tabular}{l l l l l l l l l l l l l}
\hline
 Name &  z & $A_{\lambda, BF}$ & $\alpha_{\lambda,BF}$ & $A_{3000, BF}$ & $\alpha_{\lambda,Noext}$  & $\alpha_{\lambda,SMC}$ & $P(\chi^2_{BF})$ & $P(\chi^2_{Noext})$ & $P(\chi^2_{SMC})$ & d.o.f.\\
\hline
SDSS J1148 &  6.40 &70\% SMC+30\%PISN2 &    -2.90 & 0.82 & -1.75 & -2.57 &0.43&  0.99&  0.53 & 14\\
SDSS J1623&  6.26 & 60\% SMC+40\% PISN2 &   -2.63&  0.92 & -1.39&-2.04&  0.44& 0.94& 0.47& 12\\  
SDSS J1048$^{\mathrm{a}}$ &  6.20 & 30\%SMC + 70\% H4 &  -2.90&  1.43 & -1.75 & -2.90&0.14 & 1.00 &  1.00 & 15\\  
ULAS 0202$^{\mathrm{a}}$ &  5.72 & 50\% SMC+50\% PISN2  &  -2.55 & 1.69 & -0.73 &-1.46& 0.40& 0.99 & 0.63 & 16\\
SDSS J1626&  5.20 & 60\% SMC+40\% PISN2   &  -2.90 & 1.64&  -1.00 &-2.36&  0.08& 1.00& 0.23 & 24\\  
SDSS J0852$^{\mathrm{a}}$ &  4.22 & 40\% SMC + 60\% PISN2&  -2.33&  2.00 & -0.50 & -1.14&0.01 & 1.00 & 0.92& 26 \\
APM0827$^{\mathrm{a}}$ &  3.90 & 40\% SMC+60\% PISN2 & -2.67&  1.94&  -0.90& -1.55& 0.00&  1.00 &0.04 & 30\\  
\hline
\end{tabular}
\end{center}
$^{\mathrm{a}}$ BAL quasars.
\end{table*}
\begin{table*}
\begin{center}
\caption{Simultaneous best-fit model parameters for reddened quasars}
\label{tabsim}
\begin{tabular}{l l l l l l}
\hline
Sample & $A_{\lambda, BF}$ & $P(\chi^2_{BF})$ & $P(\chi^2_{Noext})$ & $P(\chi^2_{SMC})$ & d.o.f.\\
\hline
ALL&60\% SMC+40\%PISN2 & 0.10&  1.00&  0.61 & 161\\
BAL&40\% SMC+60\% PISN2 &   0.17& 1.00& 0.94& 99\\  
noBAL& 70\%SMC + 30\% PISN2&  0.02 & 1.00 &  0.08 & 58\\  
\hline
\end{tabular}
\end{center}
\end{table*}
\begin{figure*}
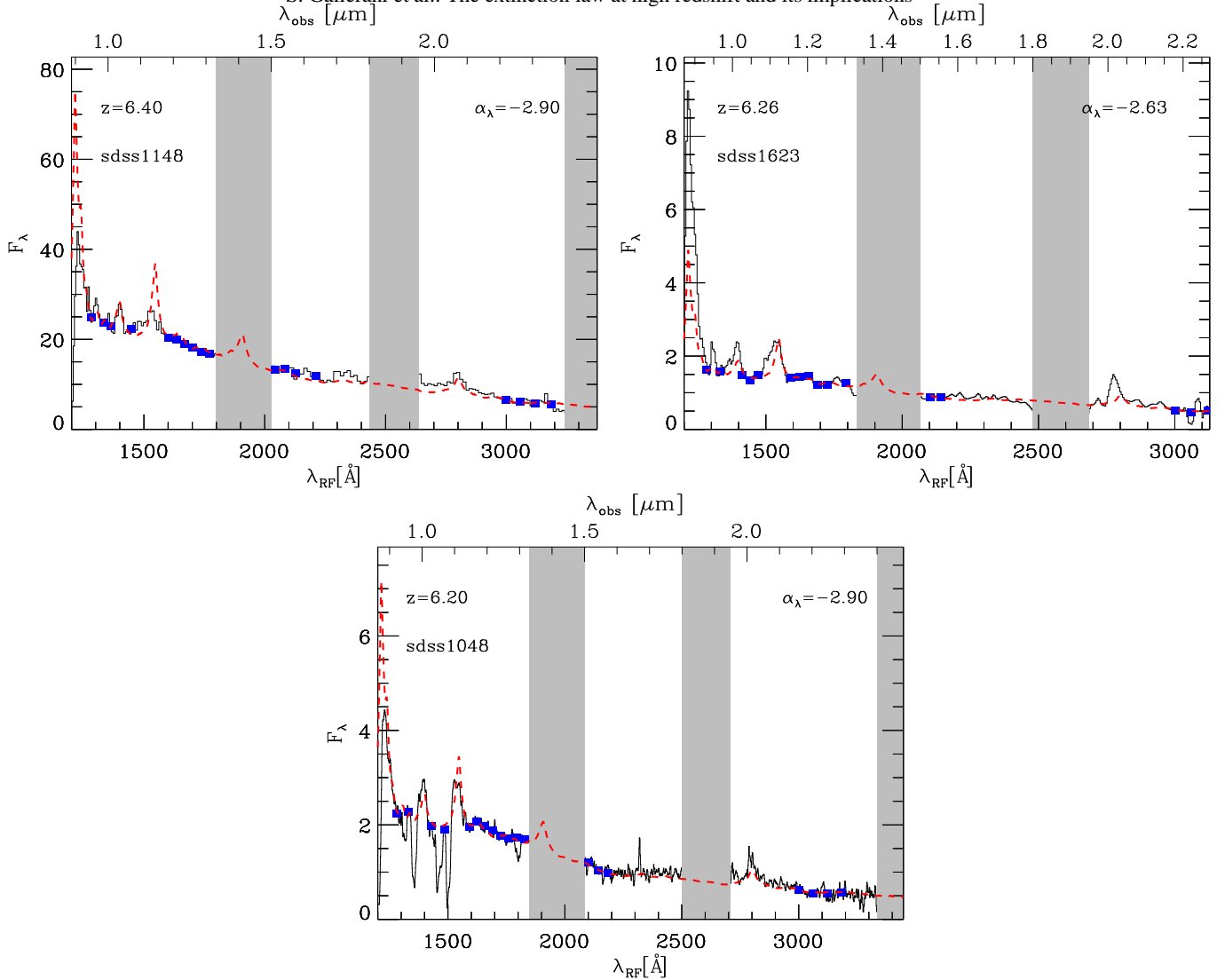

\centerline{
\psfig{figure=14721fg5.ps,width=9.5cm,angle=0}
$\!\!\!\!\!\!\!\!\!\!\!$
\psfig{figure=14721fg6.ps,width=9.5cm,angle=0}
}
\vspace{0.5cm}
\centerline{
\psfig{figure=14721fg7.ps,width=9.5cm,angle=0}
}	
\caption{Observed/rebinned spectra are indicated by black lines/filled blue squares, while red dashed lines represent the best fit spectra obtained through the $\chi^2$ analysis. The gray bands report regions of poor atmospheric transmission.
}
\label{red1}
\end{figure*}
\begin{figure*}
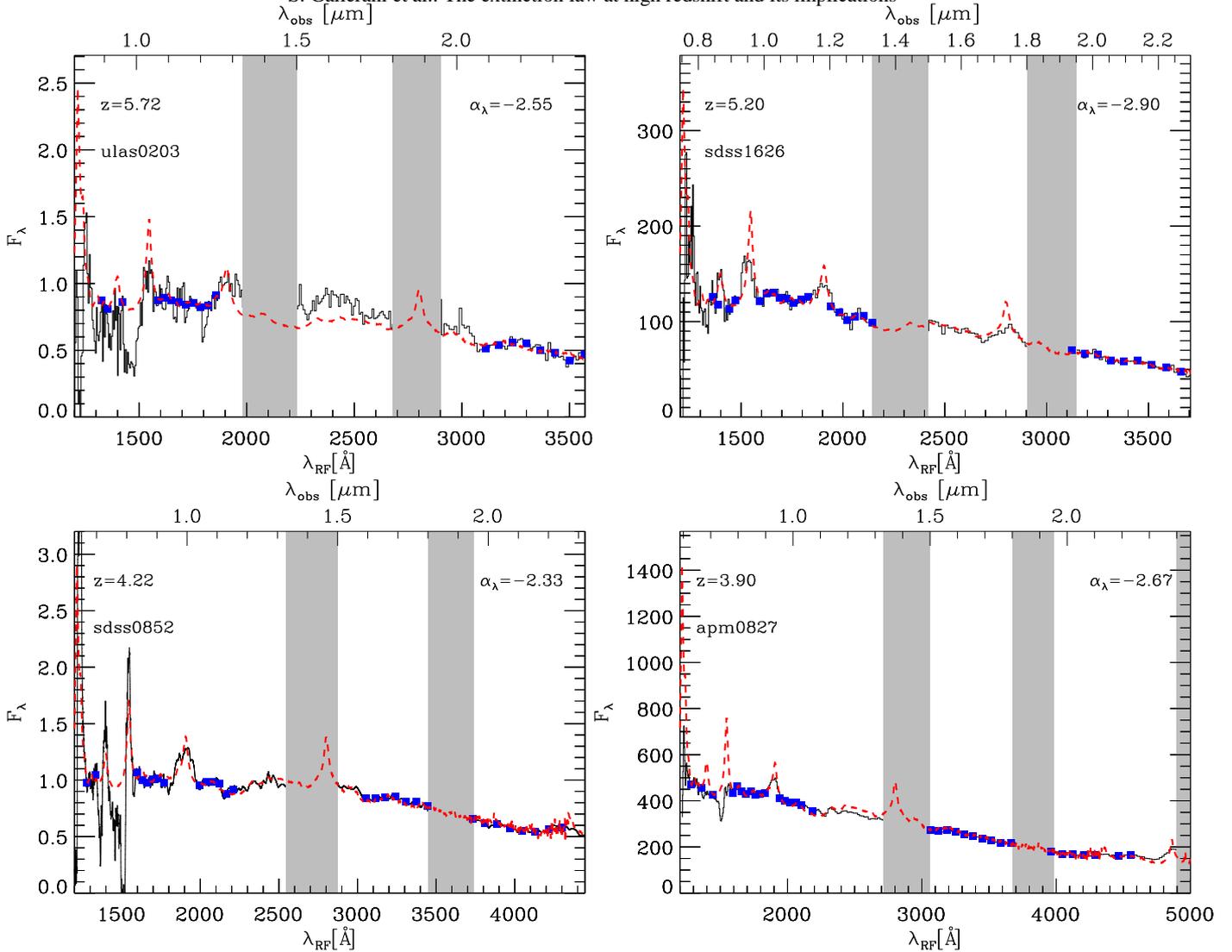

\centerline{
\psfig{figure=14721fg8.ps,width=9.5cm,angle=0}
$\!\!\!\!\!\!\!\!$
\psfig{figure=14721fg9.ps,width=9.5cm,angle=0}
}
\vspace{0.5cm}
\centerline{
\psfig{figure=14721fg10.ps,width=9.5cm,angle=0}
$\!\!\!\!\!\!\!\!$
\psfig{figure=14721fg11.ps,width=9.5cm,angle=0}
}
\caption{Same as Fig. \ref{red1}. We note a prominent FeII bump in the ULAS J0203+0012 spectrum.}
\label{red2}
\end{figure*}    

Finally, we note that the dust that causes the reddening of a quasar spectrum may not be produced by a single extinction curve among those discussed above, but by a mixture of dust grains with different
properties, origins, and extinction curves. Therefore, besides the grid of extinction curves mentioned
above, we
also consider mixed extinction curves. In particular, since the dust extinction inferred from
quasar spectra at $0<z<2$ can be described in terms of the SMC extinction curve (Richards et al. 2003;
Hopkins et al. 2004), we only mix the modeled extinction curves with the SMC, which
is our low-$z$ reference for quasars. For example, in the case of zero-metallicity extinction curve predicted by
\cite{tf01} for a SNe-like dust, we consider the following family of extinction curves   
\begin{equation}
A_{\lambda}^{mix}=(1-p)~A_{\lambda}^{SMC}+p~A_{\lambda}^{TF1},    
\end{equation}
where $\rm 0\le p\le 1$ defines the fraction of each SNe-type contributing to the extinction curve.
We consider variations in the $p$ parameter with steps of $\Delta p = 0.1$.

As far as this work is concerned, we take into account empirical and theoretical extinction curves, while we do not consider
any kind of parametric curve (e.g. Li \& Liang 2008). We found that the use of parametric curves is
hampered by a strong degeneracy among the various free parameters, 
normalization constant $C$, the continuum slope $\alpha_{\lambda}$, and the absolute extinction $A_{3000}$,
which enter in Eq. 1. We note that this problem is alleviated in the case of the GRB afterglow
spectra, for which the intrinsic slope of the continuum can be inferred from X-ray observations, which are not affected by
dust.

\subsection{Extinction curves of individual quasars}\label{redqso}

We find that seven quasars require significant dust extinction.
We note that the large fraction of extincted quasars found by ourselves ($\sim 23\%$) is much
higher than the one found in lower redshift samples ($\sim 5\%$, Hopkins et at. 2004, Reichards et al. 2003).
However, this fraction does not reflect the true fraction of reddened quasars at high redshift, being simply a selection effect, because we preferentially selected BAL quasars, which are generally
affected by higher extinction than non-BAL quasars.

In Table \ref{tab2s}, we report the quasars that
according to our analysis are reddened, along with the extinction curve minimizing the $\chi^2$, described in
Sect.~\ref{secec}, the best-fit model parameters $\alpha_{\lambda,BF}$ and $A_{3000,BF}$, the slope in the ``no extinction'' case
$\alpha_{\lambda,Noext}$, the probability associated with the best-fit values $P(\chi^2_{BF})$, the ``no
extinction'' case $P(\chi^2_{Noext})$, the ``pure SMC'' case $P(\chi^2_{SMC})$, and the degrees of freedom (d.o.f.)
associated with each spectrum. 

\begin{figure*}
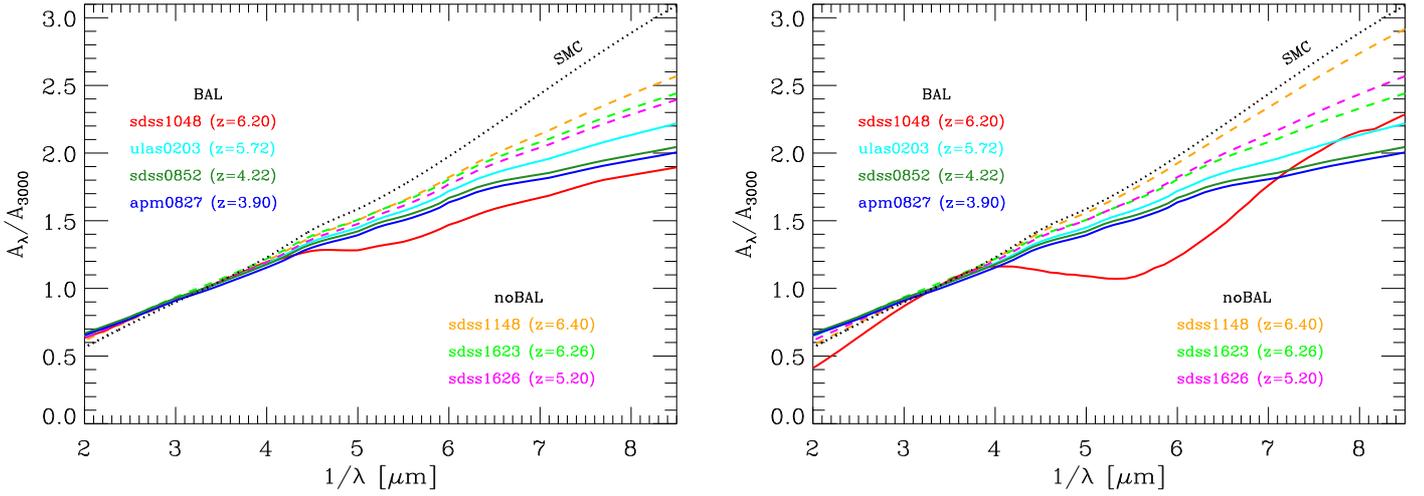

\centerline{
\psfig{figure=14721fg12.ps,width=9.5cm,angle=0}
\psfig{figure=14721fg13.ps,width=9.5cm,angle=0} 
}
\caption{
Best fit extinction curves of reddened quasars.
The solid lines are for BAL quasars, while dashed lines are for non-BAL quasars.
 For comparison, the SMC extinction curve is also shown and labeled in the figure (dotted black line).
 The panel on the left shows the results assuming a minimum intrinsic slope $\alpha _{\lambda,min}=-2.9$, while the panel
 on the right is obtained with $\alpha _{\lambda,min}=-2.6$.
}
\label{extind}
\end{figure*}

In Figs. \ref{red1}-\ref{red2}, we show the observed/rebinned
spectra (black solid line/blue filled squares), and the best-fit spectra (red dashed line) for these reddened quasars. 
Figure \ref{extind} (left panel) shows the resulting
best-fit extinction curves for each reddened spectrum (BAL quasars with solid lines and non-BAL quasars with a
dashed lines).

The inferred extinction curves for all these reddened quasars at z$>$4 deviate from the SMC, with the tendency
to flatten at short wavelengths ($\lambda_{RF}\lesssim 2000$~\AA) relative to the SMC. However, the SMC extinction curve, although not providing
a best fit for any object, is still marginally consistent with the data at the
1$\sigma$ level in most of the reddened quasars. Only for SDSSJ1048 and SDSSJ0852 is the SMC extinction curve ruled out with high confidence.

Finally, we note that
three of the seven reddened quasars require an extreme slope of $\alpha = -2.9$.
While in Sect. \ref{seleff} we discuss selection effect that probably favor the selection of high-z reddened quasars
with very blue slopes, it is worth checking the effect of restricting the allowed spectral slopes to a more
``reasonable'' range, more specifically $\alpha > -2.6$, which still characterize 90\% of quasars at lower
redshifts. The result of changing the maximum slope is shown in Fig.~\ref{extind} (right hand panel). Only for SDSSJ1048 does the shape of the extinction curve change, by preferring TF1 (as in Maiolino et al. 2004a), but the differences from previous cases are not large. We also restricted the spectral slope range to [-2.0; -1.2], which is the interval encompassed by 68\% of quasars at lower redshifts, finding that the average extinction curve (see next section) does not change significantly.

\subsection{The average extinction curve at high redshift and the deviation from the SMC}\label{MECsec}

To determine how significantly the extinction curve deviates from that of the SMC at z$>$3.9, and provide an average description of the extinction curve at high redshift, we follow two approaches.

\begin{figure*}
\centerline{
\psfig{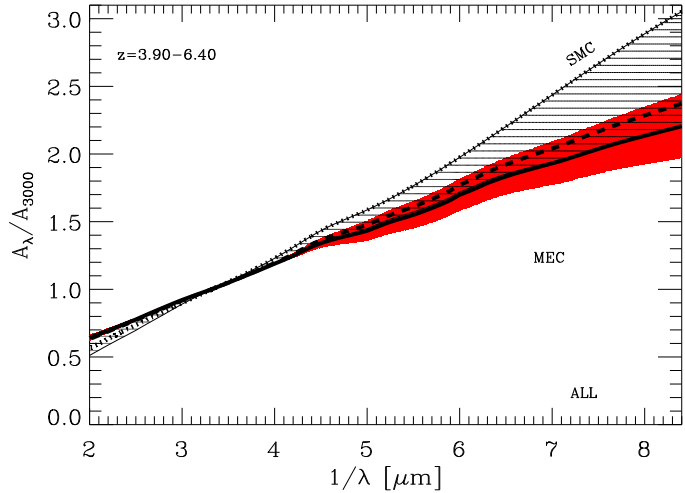} 
}
\caption{The solid line shows the mean extinction curve (MEC) computed by averaging the results obtained for all reddened quasars,
while the shaded region shows the dispersion. The dashed line shows the extinction curve obtained by the simultaneous/global fit of
all the reddened quasars (GEC), while the hatched area shows the 68\% confidence limit.
}
\label{MEC}
\end{figure*}

We first compute the mean of the inferred extinction curves to provide an empirical extinction law at $z\gtrsim 3.9$. 
The result is shown in Fig. \ref{MEC}, where the black solid line represents the mean of the inferred extinction curves. 
Hereafter, we refer to this extinction curve as the ``mean extinction curve'' (MEC). The shaded area illustrates the dispersion in the extinction curves.
It is evident that the MEC deviates substantially from the SMC, displaying a flattening at short wavelengths ($\lambda_{RF}\lesssim 2000$~\AA), although the dispersion is
high.\\

\begin{figure*}
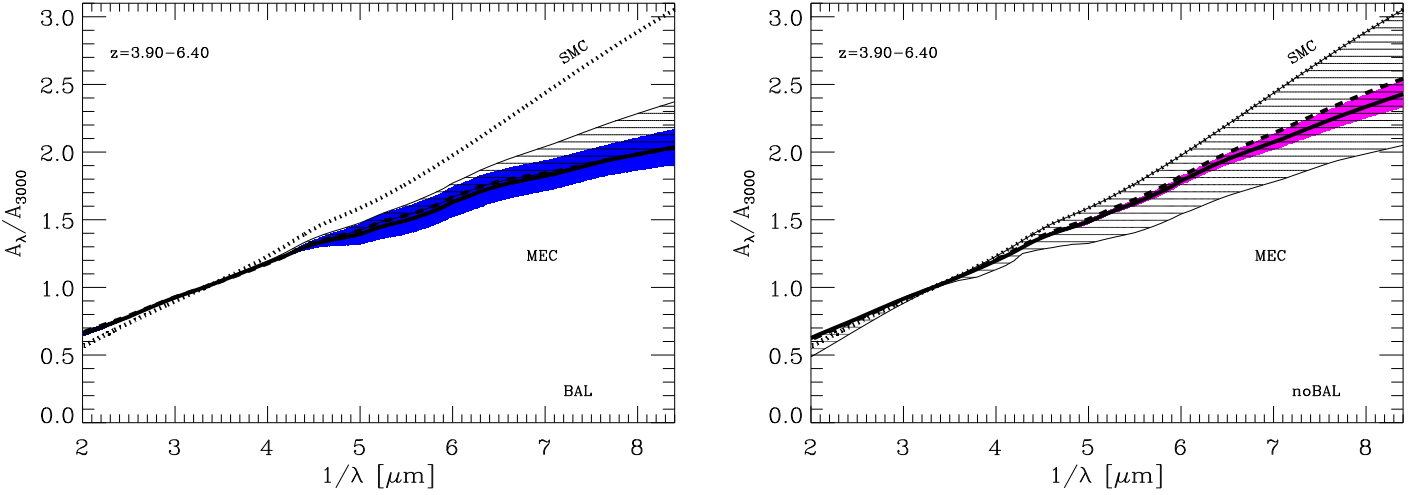

\centerline{
\psfig{figure=14721fg15.ps,width=9.5cm,angle=0} 
\psfig{figure=14721fg16.ps,width=9.5cm,angle=0} 
}
\caption{Mean (MEC) and simultaneous/global (GEC) extinction curve of reddened quasars divided into BAL (left) and non-BAL (right).
The coding is the same as in Fig.~\ref{MEC}.
}
\label{mectype}
\end{figure*}

Secondly, we perform the simultaneous fit of all the seven reddened quasars; we report the results of this second approach in Table \ref{tabsim}. The resulting global best-fit extinction curve (GEC) is indicated by a black dashed line in Fig.~\ref{MEC}, and is very close to the MEC. In this case, we can also infer the confidence limits, which are similar to those obtained for individual quasars, as shown by the hatched area in Fig.~\ref{MEC}.
Although the simultaneous fitting should improve and decrease the width of the confidence limits (thanks to the higher quality statistics than for individual spectra), the confidence interval remains large implying that the curve of the SMC is consistent 
with the GEC. As we show in the following, most of the dispersion and the width of the confidence intervals is due to
the non-BAL quasars.
Originally, the difference between BAL and non-BAL quasars was mostly ascribed to different viewing angles of objects that
are intrinsically similar \citep{weymann91,goodrich95,hines95}.
However, the finding of systemic differences between BAL and non-BAL quasars in terms of observational properties that should be isotropic, has suggested that these two classes of quasars are intrinsically
different. In particular, the finding that the radio morphology of BAL is generally
more compact than those of non-BAL quasars suggests that the former are in an earlier evolutionary stage \citep{becker00}.
If BAL and non-BAL quasars were to trace different evolutionary stages, this could also imply different dust properties.
Different extinction curve properties are already apparent in Fig.~\ref{extind}: the extinction curves of
BAL quasars (solid curves) are systematically flatter than those of non-BAL quasars (dashed
curves).
In Fig.~\ref{mectype},
we show the MEC and GEC for BAL and non-BAL quasars separately. The average extinction curve (MEC)
of BAL quasars exhibits a much lower dispersion than the whole sample of reddened quasar. Moreover, for this subsample
the confidence intervals on the simultaneous/global fit (GEC) are much narrower. As a consequence, for BAL quasars,
the deviation of the simultaneous/global extinction
curve from the SMC is far more significant than in the case of the whole sample of quasars; in particular,
the deviation from the SMC is significant at a confidence level higher than 95\%.
The MEC and GEC of non-BAL quasars are intermediate between the extinction curve of BAL quasars and the SMC.
The confidence intervals of the GEC for non-BAL quasars are much wider, making the GEC consistent with the SMC curve.

Our speculative interpretation is that the extinction curve of high-z BAL quasars reflects the dust properties
in the early evolutionary stages of their host galaxies, when SNe possibly provided a preferential mechanism for dust
production. High-z non-BAL quasars are probably in a more evolved
stage, characterized by dust properties that are in-between those of the SMC and those of primordial dust grains.
However, we cannot exclude that the different extinction curve of high-z BAL quasars is related to a different reprocessing
of dust grains in the ISM, from those of lower redshift and non-BAL quasars. For instance, higher densities may favor the growth and coagulation
into larger grains flattening the extinction curve. The preferential destruction of small grains due to sublimation is also important; indeed the sublimation radius for small grains is larger than that of large grains \citep{laordraine93}. At high quasar luminosities, such as those characterizing high-$z$ quasars (because of selection effect) this differential sublimation effect may play a major role. If the latter is the main explanation of our findings, then the extinction curve depends mainly not on redshift but on quasar luminosities. In this case, the flattening is only associated with high-$z$ quasars and not with early galaxies.  

The higher confidence with which the GEC of BAL quasar deviates from the SMC is also most likely caused by observational effects
and in particular by BAL quasars being generally more reddened, and their extinction curve therefore being able to be measured with higher accuracy.

We finally emphasize that our result is based on the analysis of individual quasars and not on the use of composite spectra,
a method that can infer biased (artificially flatter) extinction laws \citep{will05}.
\subsection{Selection effects and spurious correlations}\label{seleff}
\begin{figure*}
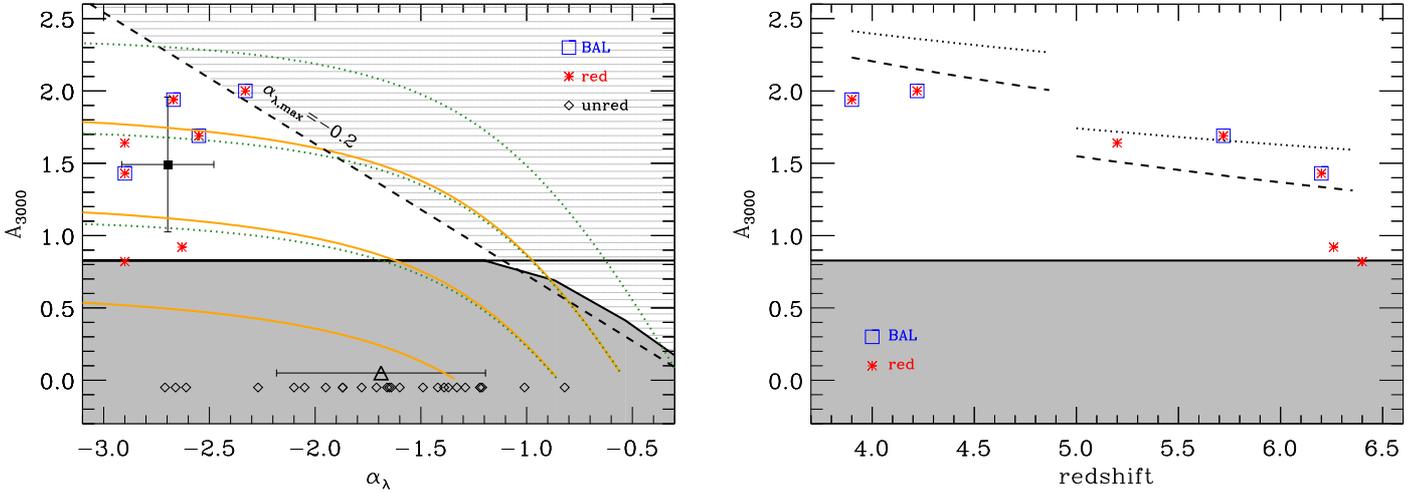

\centerline{
\psfig{figure=14721fg17.ps,width=9.5cm,angle=0}
\psfig{figure=14721fg18.ps,width=9.5cm,angle=0}
}
\caption{{\it Left hand panel}: Red asterisks show the best fitting values of $A_{3000}$ and $\alpha_{\lambda}$, as obtained for reddened quasars.
Red asterisks surrounded by blue squares represent BAL quasars. The black filled square represents the corresponding mean value, and the error
bars the standard deviations. The empty black triangle and the associated error bars denote the $\alpha_{\lambda}$ mean value and the
$1\sigma$ dispersion in the case of unreddened quasars. Solid orange and dotted green curves represent the maximum value of $A_{3000}$, such
that, for a given bolometric luminosity $L_{bol}$, the observed flux at a given wavelength is greater than the flux detection limit in the
Z-band and I-band, respectively. 
The bolometric luminosity has been varied in the range $10^{47}<L_{bol}<10^{48}$ (curves from the bottom to the top). The orizontal black solid line represents the minimum extinction required for a quasar to be classified as ``dust reddened''. Quasar spectra reddened by an extinction lower than this minimum value (gray shaded area) can still be fitted with a power law redder than the intrinsic one, and classified as ``unreddened''. The black dashed line shows the effect of quasar color selection criteria on our results. High-z quasars are selected to have a blue color on the filters redwards of the dropout (z--J and i--z). This requirement can be translated into a maximum {\it observed} spectral slope ($\rm \alpha _{\lambda, max}$) which implies
a maximum extinction $\rm A_{3000,max}$ affecting quasars. Clearly $\rm A_{3000,max}$ depends on the intrinsic slope $\alpha _{\lambda}$ of the quasar,
and it is shown with the black dashed line for $\rm \alpha _{\lambda, max}=-0.2$, i.e. the minimum of the slope distributions in the SDSS quasar survey. Quasars reddened by an extinction higher than the dashed line (gray hatched region) are lost by the quasar color selection criteria. {\it Right hand panel}: Trend of $A_{3000}$ with redshift. The lines represent the maximum value of $A_{3000}$ for which a reddened spectrum, for a fixed bolometric luminosity ($L_{bol}=10^{48}~\rm{erg~s^{-1}}$), is still detectable by the SDSS. The results have been obtained assuming a slope for the intrinsic continuum $\alpha_{\lambda}=-2.4$ (dotted), $\alpha_{\lambda}=-1.6$ (dashed). The horizontal black solid line and the gray shaded region have the same meaning of the ones in the left hand panel.}
\label{A3alpha}
\end{figure*} 

Fig.~\ref{A3alpha} (left-hand panel) shows the extinction $\rm A_{3000}$ as a function of the spectral index $\alpha _{\lambda}$.
The extinction of quasars that are not reddened according to our criteria is set to be zero. The empty triangle shows the
average slope of ``unreddened'' quasars, $\langle \alpha _{\lambda} \rangle = -1.68$, which is fully consistent with the 
average slope of unreddened quasars at low-intermediate redshifts (Reichard et al. 2003). This indicates that there is
little or no evolution in the intrinsic spectrum of quasars at high-z and also supports our assumption
that low-z templates are appropriate for modeling high-z quasars.

However, we note that all quasars identified as ``reddened'' are characterized by very blue slopes, on the
tail of low-intermediate quasar distribution. This is most likely the combined consequence of selection effects and our (conservative) criteria for identifying reddened quasars, as discussed in the following. A reddened power law can still
be fitted with a (redder) power law, and classified as ``unreddened'', as long as the extinction is not too high.
The deviations from a power law become significant only beyond a ``minimum'' extinction. For each intrinsic $\alpha _{\lambda}$, we inferred the minimum $\rm A_{3000,min}$ for which the deviations from a (redder) power law become significant and cause the target to be classified as ``dust reddened''. We plot $\rm A_{3000,min}$ in Fig.~\ref{A3alpha} with a thick, black solid line,
which has a value of about 0.8~mag and is independent of the intrinsic $\rm \alpha_{\lambda}$.
Quasars affected by extinction lower than this minimum
value (gray shaded area) are not classified as ``absorbed'' according to our criteria. That $\rm A_{3000,min}$
decreases at $\alpha _{\lambda} >- 1.2$ is simply a consequence of our limited range of intrinsic slopes ([-2.9; -0.2], Sect. \ref{method}),
which makes an absorbed quasar spectrum intrinsically red and difficult to fit with a redder slope, simply because of the ``lack'' of redder
slopes in the template library. If we remove the lower limit to the allowed intrinsic slopes, then we obtain the horizontal solid thick line,
which extends the  $\rm A_{3000,min}\sim 0.8$ line to $\rm \alpha _{\lambda}=-0.3$.

Another selection effect is related to the parent sample of high-z quasars, which were selected as color dropouts (i--z at z$\sim$6 and
r--i at z$\sim$4.5, which sample the IGM Ly$\alpha$ forest break at the respective redshift), and to also have a blue color on the filters rewards
of the dropouts (z--J and i--z). The latter selection criterion is required in order to minimize the contamination by brown dwarfs and low redshift interlopers. The latter requirement
can be translated into a maximum {\it observed} spectral slope ($\rm \alpha _{\lambda, max}$). This implies
a maximum extinction $\rm A_{3000,max}$ affecting quasars, which corresponds to the reddening that causes the {\it observed} slope to be redder than $\rm \alpha _{\lambda, max}$. Clearly $\rm A_{3000,max}$ depends on the intrinsic slope $\alpha _{\lambda}$ of the quasar,
and it is shown with a dashed line in Fig.~\ref{A3alpha}, where we have assumed $\rm \alpha _{\lambda, max}=-0.2$, as inferred
by the minimum of the slope distributions in the SDSS quasar survey. Quasars reddened by an extinction higher than
the dashed line (gray hatched region) are lost by the quasar color selection criteria.

Quasars surviving both our definition of reddened quasars (above the gray shaded area) and the survey
color-selection criteria (below the gray hatched area) are clearly limited to a small region of the $\rm A_{3000}-\alpha _{\lambda}$
domain, mostly in the low $\alpha _{\lambda}$ range, which is indeed where our reddened quasars are identified.

Another selection effect is linked to the detectability of quasars as a function of spectral slope and extinction.
For a given bolometric luminosity, quasars with a bluer slope are easier to detect because a higher fraction of the flux
is emitted in the UV, which is the band where quasars are detected. As a consequence, for a given bolometric luminosity,
dust extinction brings intrinsically red quasars below the survey detection threshold more easily than blue quasars. This effect
is illustrated by the dotted lines in Fig.~\ref{A3alpha}, which show the maximum $\rm A_{3000}$ at which a quasar with a given bolometric
luminosity can still be detected by the limiting magnitude of the survey in the i-band ($\rm z\sim 4$, green lines) or 
in the z-band ($\rm z\sim 6$, orange lines) as a function of the spectrum slope. For each of the two cases, the three lines
correspond to different luminosities ($\rm log~L_{bol}= 47, 47.5, 48~erg~s^{-1}$). This selection effect clearly also contributes
to bias the sample of reddened quasars against intrinsically red objects. However, the true magnitude of this effect is difficult
to quantify because of claims of a possible correlation between luminosity and spectral slope in quasars
(e.g. Pu et al. 2006). As a consequence, the dependence of the maximum $\rm A_{3000}$ on $\alpha_{\lambda}$ may be globally steeper
for the real population of high-z quasars than illustrated by the dotted lines in the left hand panel of Fig.~\ref{A3alpha}.

Figure ~\ref{A3alpha} (right hand panel) shows the extinction $\rm A_{3000}$ as a function of redshift for reddened quasars, showing an anticorrelation
between these two quantities. However, this anticorrelation is also probably mostly driven by selection effects: for a given
bolometric luminosity, more distant quasars appear fainter and therefore fall below the detection threshold of the survey with a lower
amount of dust extinction. To illustrate this effect, the dashed and dotted lines in Fig.~\ref{A3alpha} (right hand panel) illustrate the maximum extinction
allowed for quasars with bolometric luminosity of $\rm 10^{48}~ergs~s^{-1}$ (with $\alpha=-2.4$ and $\alpha=-1.6$, respectively) in order
not to be missed by the SDSS survey, which approximately reproduces the trend observed in our reddened quasars.
The ``jump'' at z$\sim$5 is caused by the different photometric bands used to select quasars in the
two redshift ranges. Our reddened quasars approximately follow the expected relation. Although our data have very limited statistics,
at very high redshift (z$>$6.3) it seems that there is a shortage of highly reddened quasars that are still expected within the
selection effects (extrapolation of the dashed and dotted curves). The latter result may be indicative of a real lack of dust at these very
early epochs because of the shortage of time to produce dust, in line with the finding of \cite{jiang10}. Another possible effect is that, since the Ly$\alpha$ cutoff of these objects is partly outside the z-band, they would need to have very high instrinsic luminosities to be reddened and observed by SDSS; since the luminosity function is very steep at the bright end, this may cause a shortage of reddened quasars at z$>6.3$ relative to their lower redshift counterparts. However, the statistical
significance of our result is still marginal. Finally, we note that there is a lack of quasars at z$<$5 with extinctions
$\rm A_{3000}$ in the range 0.8--1.8 mag, which should be in principle observed. The statistics are low enough to ensure that this shortage of quasars with intermediate extinction is not a real worry. However, we note that at z$<$5 many quasars are only observed with
FORS2, hence their data have a very limited spectral coverage; as a consequence, for these quasars the minimum extinction required to be
classified as ``reddened'' is significantly higher, approaching $\rm A_{3000,min}\sim 1.5~mag$, hence partially explaining
the additional selection effect noted in Fig.~\ref{A3alpha}.
\subsection{Degeneracy between extinction curves}

\begin{figure}
\centerline{
\psfig{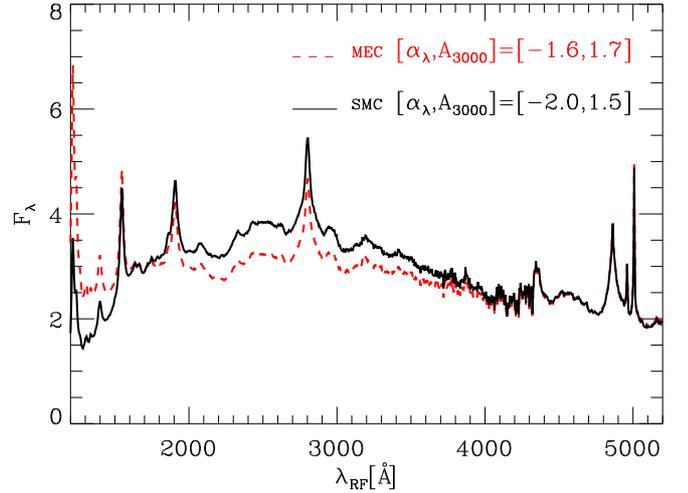}
}
\caption{Synthetic spectra obtained by assuming a SMC-like reddening with $[\alpha_{\lambda}, A_{3000}]=[-2.0,1.5]$ (solid black line)
and a MEC-like reddening with $[\alpha_{\lambda}, A_{3000}]=[-1.6,1.7]$ (red dashed line).} 
\label{deg}
\end{figure}

An SMC-extincted spectrum with a given slope is very similar to an intrinsically redder spectrum absorbed by an extinction curve
that, relative to the SMC, flattens in the UV. This effect is illustrated in Fig.~\ref{deg}, which shows
two spectra with intrinsic slopes of $\alpha_{\lambda}=-1.6$ and $\alpha_{\lambda}=-2.0$, but absorbed with the MEC and SMC extinction
curves, respectively. The two resulting spectra are indeed similar, although we verified by means of simulations
that with our analysis we can recover the correct extinction curve in both cases. However, what is interesting within this context,
is that any bias introduced by this degeneracy goes in the opposite direction of the results obtained by us.
Indeed, that we find a deviation from the SMC extinction curve flattening in the UV (the MEC), and a blue intrinsic slope (as discussed in the previous section) strengthens our finding,
meaning that in an unbiased quasar sample (i.e. with a lower limiting flux, hence with an intrinsic slope closer to the average of unreddened
quasars) the deviation from the SMC may be even more pronounced than the conservative result reported here.

We finally note that there is an unavoidable degeneracy in the {\it slope} of the extinction curve, which is intrinsic in {\it any}
study using only the spectral shape (i.e. without knowledge about the intrinsic luminosity of the object). Given an extinction
curve $\rm f_\lambda = A_\lambda / A_{3000}$, fitting an observed spectral shape, any other extinction curve of the form
$\rm f'_\lambda = k(f_{\lambda}-1)+1$, can indeed fit equally well the same observed spectrum with any value of $\rm k$, by simply changing
the absolute extinction $\rm A_{3000}$. We reiterate that this degeneracy affects {\it any} study attempting to infer extinction
curves solely based on the objects spectral shape (i.e. the vast majority of previous studies). Only in cases where the intrinsic
luminosity is known, and therefore $\rm A_{3000}$ is not free to vary, can the slope ($\rm k$) of the extinction law be determined.
The alternative approach, as adopted here, is to trust only a limited library of (physically justified) extinction curves,
without leaving the possibility of changing their individual slope.

\section{Implications for primordial galaxies}
\subsection{Observational implications}

We have inferred the extinction curve for quasars at z$>$4. We propose that the same extinction curve applies also to
star-forming galaxies at high-z in general and adopt this as a working assumption. 
If the dust properties at high redshift differed significantly from those at low redshift, this would have important
effects for the observational properties of high-$z$ galaxies. The extinction curve adopted to
correct the rest-frame UV radiation of high-$z$ galaxies
should be modified with respect to low redshift templates, otherwise the
inferred intrinsic luminosity and spectral shape (used to infer the star-formation rate and stellar population)
will be erroneously estimated. In particular, the evolution of the cosmic star-formation rate at $z>4$
may be erroneously traced if the change in the dust extinction properties is not properly accounted for.
In this section, we discuss, and when possible quantify, these observational implications.

\subsubsection{Attenuation curves}
\begin{figure*}
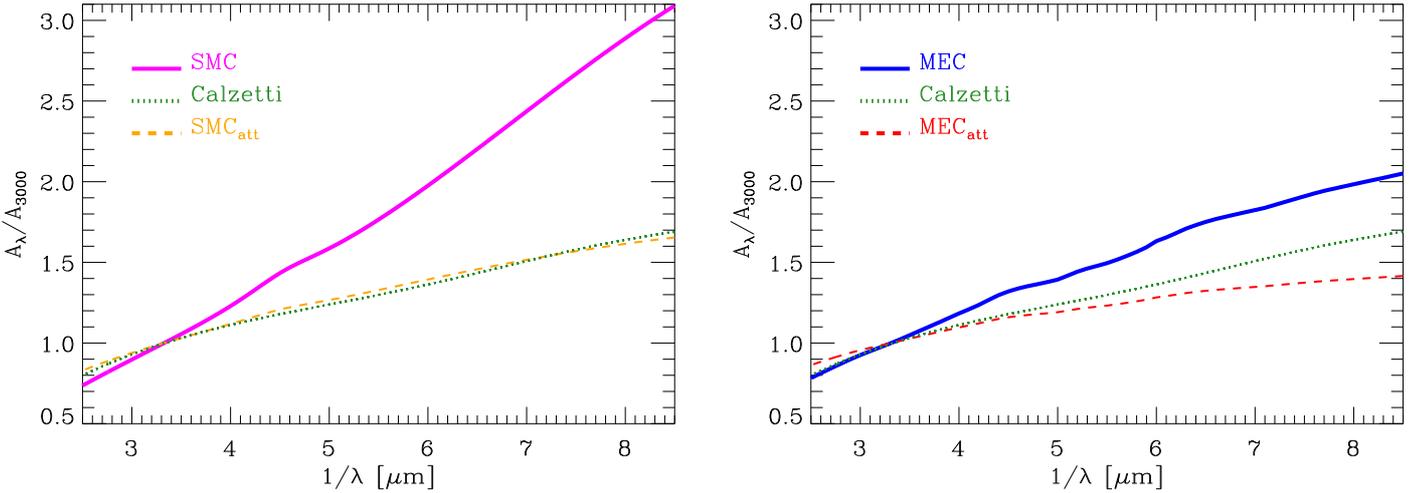

\centerline{
\psfig{figure=14721fg20.ps,width=9.5cm,angle=0}
\psfig{figure=14721fg21.ps,width=9.5cm,angle=0}
}
\caption{{\it Left hand panel}: The attenuation curve corresponding to the SMC extinction curve (solid magenta line) is shown by an orange dashed line. For comparison, the Calzetti law is also indicated by a dotted green line. It has been shown that the Calzetti law is the result of the attenuation by a medium with an SMC-type dust \citep{gordon07,inoue05}. We can recover this result using Eq. 3. {\it Right hand panel}: Same as in the left hand panel, but for the MEC. The attenuation curve resulting from the MEC is flatter than the Calzetti law, which is generally adopted for quantifying the reddening of galaxies at all epochs.}
\label{attenuation}
\end{figure*} 
For quasars, the simple dusty ``screen'' geometry applies, but for galaxies one has to consider that dust is
mixed with the emitting sources (either stars or ionized gas). As a consequence, for galaxies it is more practical
to use ``attenuation curves'', i.e. the ratio of the observed spectrum to the intrinsic total light emitted by the whole system,
before dust absorption. The attenuation curve is of course associated with the extinction curve (which is intrinsic to the
dust properties), but also depends on the geometry of the emitting sources relative to the dusty medium, as well as the viewing angle. We attempt to derive a simple version of the attenuation curve associated with the MEC
extinction curve, which can be applied to high-z galaxies. However, we also consider the simple ``screen'' case, since this
geometry seems to apply to some high-z galaxies \citep{reddy10}.

For galaxies in the local Universe, the effect of dust is ascribed to the attenuation law by Calzetti et al. (1994). It has been
shown that the Calzetti law is the result of the attenuation by a medium with an SMC-type dust \citep{gordon07,inoue05}.
Therefore, that the MEC is flatter in the UV
relative to the SMC indicates that the attenuation law resulting from a MEC-type dust must be
flatter than the Calzetti one. We assume a very simple case where dust and stars are uniformly mixed and the density is constant throughout the galaxy. In this case, it is easy to infer that the attenuation curve normalized to 3000~\AA \ is given by

\begin{equation}\label{att}
\frac{A_{\lambda}^{att}}{A_{3000}^{att}}=\frac{log_{10}\left(\frac{1-e^{-\frac{A_\lambda}{A_{3000}}\tau _{3000}}}{\frac{A_{\lambda}}{A_{3000}}\tau_{3000}}\right)}
{log_{10}\left(\frac{1-e^{-\tau _{3000}}}{\tau_{3000}}\right)},
\end{equation}
where $\rm A_\lambda /A_{3000}$ is the dust extinction curve, and $\tau _{3000}$ is the {\it total} dust optical depth at 3000~\AA \ of the
dust throughout the galaxy (i.e. the dust optical depth integrated over the entire thickness of the galaxy along the line of sight).
First, we insert in Eq. (\ref{att}) the SMC extinction curve, and we show in Fig. \ref{attenuation} that we can almost perfectly
reproduce
the Calzetti law with $\tau_{3000}=6$. We then investigate the effect of changing the underlying extinction
curve by keeping the same value of $\tau_{3000}$ and using the MEC instead of the SMC; we refer to the resulting attenuation
curve with the notation MEC$_{att}$. The result is shown in the right-hand panel of Fig.~\ref{attenuation},
where the blue solid line denotes the MEC, the dashed red line the MEC$_{att}$, and the green dotted line the Calzetti law, which is reported for comparison. We note that for this analysis we use the MEC obtained from the analysis of BAL quasars, which (as discussed in Sect. \ref{MECsec}) we consider as the truly ``primordial'' extinction curve.
We indeed confirm our expectations, i.e. that the attenuation curve inferred by the MEC is flatter than the Calzetti one.

\subsubsection{Reddened galaxy spectra}
\begin{figure*}
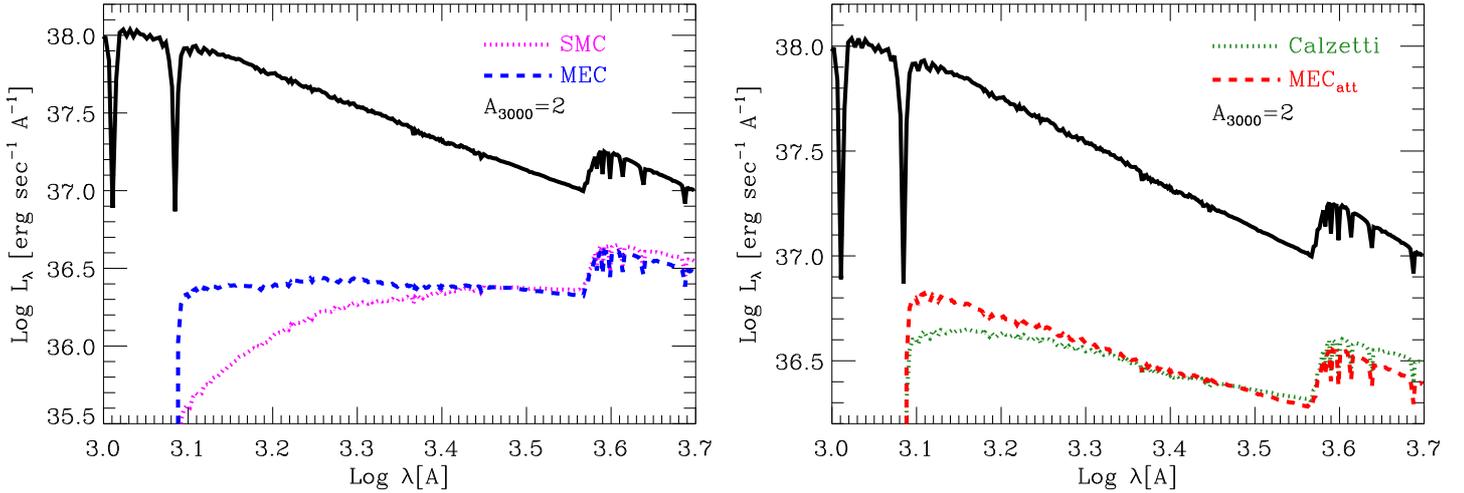

\centerline{
\psfig{figure=14721fg22.ps,width=9.5cm,angle=0}
\psfig{figure=14721fg23.ps,width=9.5cm,angle=0}
}
\caption{Spectra of galaxies at redshifts higher than the reionization epoch. The solid black lines represent unabsorbed templates for a galaxy $70$ Myr old, having a metallicity $Z=5\times10^{-2} Z_{\odot}$, which has experienced an instantaneous star-formation, resulting in the production of  $10^6$ M$_{\odot}$ of stars. Dotted and dashed lines show the effects of different types of reddening, as labeled in the figure. The spectral region blueward of the Ly$\alpha$ is assumed to be completely absorbed by the intervening intergalactic medium.}
\label{gal1}
\end{figure*} 
We now investigate how different extinction and attenuation curves affect the shape of a typical galaxy spectrum.
As an example, we consider a 
(unabsorbed) spectrum of a STARBURST99 model \citep{leitherer99} for a galaxy $70$ Myr old that has experienced
an instantaneous burst of star-formation, resulting in the production of
$10^6$ M$_{\odot}$ of stars. We assume a metallicity of $Z=5\times 10^{-2}$Z$_{\odot}$ and apply the constraint that the spectral region blueward of the Ly$\alpha$ is completely absorbed by the intervening intergalactic medium (which is appropriate for
galaxies at redshifts approaching the reionization epoch). Since we only aim to make an illustrative comparison between the extinction effects of SMC and MEC, we neglect the effect of nebular lines,
although they are certainly prominent in early star-forming galaxies, especially at low metallicity \citep{schaerer10}.

For the case of pure screen extinction, in the left-hand panel of Fig.~\ref{gal1}, we compare the effect of the SMC-like (magenta dotted line) and MEC-like (blue dashed line) reddening
on the spectrum of the high-$z$ star-forming galaxy template (black solid line), by assuming the same extinction
at 3000~\AA \ ($\rm A_{3000}=2~mag$). 

The MEC clearly makes the spectrum bluer and more luminous than the SMC, as expected.
This first result is particularly propitious for the knowledge of the high-$z$ Universe, since it implies a much easier detection
of early galaxies with future facilities (e.g. JWST, ELT).
In the specific case considered here, there is a difference of
nearly a factor of ten between continuum intensity at the Ly-break between the use of the two different extinction curves. What is most relevant
is that if the apparently blue spectrum produced by the MEC extinction is interpreted using the SMC extinction curve,
then one infers that the dust extinction is much
lower than the one actually affecting the spectrum.

In the right-hand panel of Fig.~\ref{gal1}, we show the reddening of galaxy spectra using the attenuation curves and, in particular,
the Calzetti law versus the MEC attenuation curve. In this case, the effect is much lower than in the case of pure screen extinction.
However, in the case of attenuation by MEC the spectrum still appears bluer than in the case of attenuation with the Calzetti law,
hence leading to an underestimate of the effect of dust if not interpreted properly.
We quantify this effect in the next section.
 
\subsubsection{Correction for dust extinction in galaxies}

\begin{figure}
\centerline{
\psfig{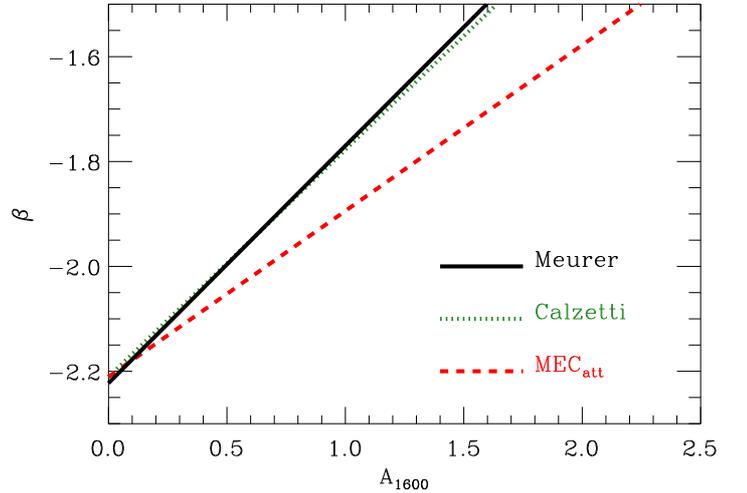}
}
\caption{Slope ($\beta$) of the UV-continuum of a galaxy as a function of the extinction/attenuation at 1600~\AA~($A_{1600}$). The solid black line show the Meurer et al. 1999 relation. The green dotted (red dashed) lines shows the $\beta$-$A_{1600}$ relation for a galaxy with an intrinsic slope of $\beta=-2.2$, assuming that the dust attenuation is described by the Calzetti law (MEC$_{att}$).}
\label{gal2}
\end{figure}

For high-z galaxies, which are generally much fainter than quasars, only a few broad-band photometric measurements
of the rest-frame UV continuum are usually available. As a consequence, the effect of dust reddening is often computed by simply
deriving the spectral slope $\beta$ ($F_{cont}\propto \lambda^{\beta}$), and measuring its deviation from an average slope assumed to represent the unabsorbed case ($\beta\sim -2.2$). The relationship between slope $\beta$ and extinction at
$\rm \lambda _{rest}=1600~\AA$ was calibrated by \cite{meurer99} and subsequently adapted by \cite{bou09} to the case of galaxies
at high-z, whose slopes are inferred by using only two broad-band filters. The relation inferred by \cite{bou09} is
\begin{equation}
\rm A_{1600} = 4.91+2.21~\beta,
\label{eq_bowens}
\end{equation}
where $\beta$ is computed in the wavelength range between 1600~\AA \ and 2300~\AA. The resulting relation is shown in Fig.~\ref{gal2} with a black solid line.
The relation between extinction and spectral slope obviously depends on the extinction or attenuation curve assumed. The relation in
Eq.~\ref{eq_bowens} can be reproduced in terms of extinction by a Calzetti attenuation curve, as illustrated by
the green dotted line in Fig.~\ref{gal2} (see also Table 2 of Meurer et al. 1999). When we adopt the attenuation
curve associated with the MEC at z$>$4, then the relation between $\rm A_{1600}$ and $\beta$ is shallower, as a consequence
of the attenuation curve being flatter in the UV than the Calzetti curve.
More specifically, in this case, the relation given by Eq.~\ref{eq_bowens} becomes
\begin{equation}
\rm A_{1600} = 9.61+4.33~\beta,
\label{eq_gallerani}
\end{equation}
where we keep the same assumption about the average intrinsic slope as in Eq.~\ref{eq_bowens}, i.e. $\beta\sim -2.2$. We recall that Eq. \ref{eq_gallerani} is obtained by adopting the MEC attenuation curve, computed based on the assumption that dust and stars are uniformly mixed and the density is constant through the galaxy.
The latter relation is compared with the one used by \cite{bou09} in Fig.~\ref{gal2}. If the reddening curve inferred
for quasars at z$>$4 also applies to high-z galaxies, then we suggest that Eq.~\ref{eq_gallerani} is more appropriate
for describing the extinction in high-z galaxies than Eq.~\ref{eq_bowens}. Both of them of course rely on the assumption
that the average intrinsic slope $\beta$ is about $-2.2$, which may not be appropriate for high-z metal poor galaxies
\citep{schaerer10}, although our main goal here is simply to show the effect of the different extinction curve
on the inferred correction of the  luminosity (and star-formation rate) in high-z galaxies.

\subsubsection{The star-formation rate at high-$z$}
\begin{figure*}
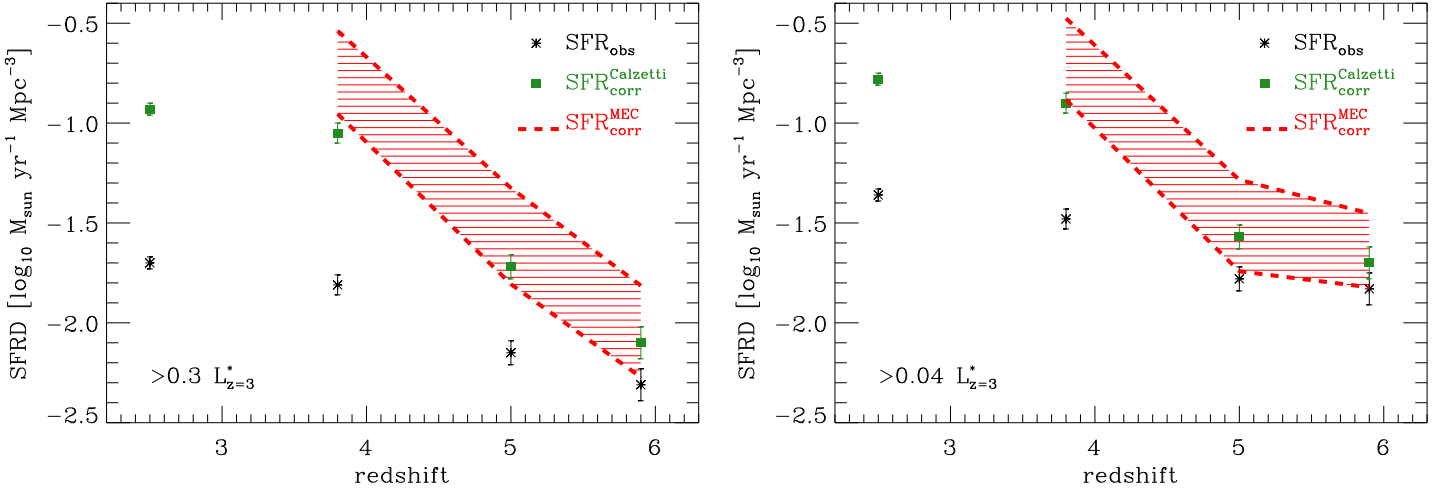

\centerline{
\psfig{figure=14721fg25.ps,width=9.5cm,angle=0}
\psfig{figure=14721fg26.ps,width=9.5cm,angle=0}
}
\caption{Effects of different dust corrections on the star-formation rate evolution for luminous (on the left) and less luminous (on the right) galaxies. Black asterisks and filled green squares represent, respectively, the observed and corrected star-formation rate (along with the corresponding errors) by Bouwens et al. (2009), assuming the $\beta-A_{1600}$ relation by Meurer et al. 1999. The red dashed lines report the corrected star-formation rate, obtained taking into account the results of this study.}
\label{madauplot}
\end{figure*} 
Dust correction is a key requirement when measuring the star-formation rate density evolution. Observations of high-$z$
galaxies are consistent with a star-formation rate density at $z\sim 6$ that is $\sim 6$ times less than that inferred
for lyman break galaxies at $\sim 3-4$ (Bunker et al. 2004; Bunker et al. 2009; Bouwens et al. 2009). At z$\sim$7, the star-formation
rate density seems to be even lower, a factor of about 10 less than the one inferred for low-z LBGs. 
Moreover, \cite{bou09} find that the UV-continuum slope of $z\sim 5-6$ galaxies is substantially bluer than $z\sim 2-4$
ones, therefore claiming that high-$z$ galaxies are less obscured by dust than their lower redshift counterparts.
However, it is intriguing that the change in slope of the continuum of galaxies occurs just at the redshift (z$\sim$4-5) where
we find that the extinction curve may change. This suggests that the change in the average continuum slope of galaxies is actually
due to a change in the dust properties responsible for reddening the spectra.

\cite{bou09} indeed mention that changes in the mean dust extinction curve may be responsible for the redshift dependence of the UV-continuum slope,
although their dust
corrections, used to infer the star-formation rate density at high-z,
are computed starting from empirical relation between $\rm A_{1600}$ and $\beta$ obtained for
galaxies in the local Universe (Eq.~\ref{eq_bowens}). In the following, we investigate the effect on the star-formation
rate density when the redshift dependence of the average continuum slope is ascribed to a variation in the dust extinction curve
at high-z.

We do not aim to perform a very detailed investigation, which would require a remodelling of all the photometric data points in galaxies
at z$\ge$4 with the MEC attenuation curve, which will instead be the topic of a separate, dedicated paper. We simply take
the star-formation rate density observationally inferred by \cite{bou09} and recompute the dust extinction corrections by
using the relation between $\rm A_{1600}$ and $\beta$ given by Eq.~\ref{eq_gallerani}, which is appropriate for the MEC attenuation curve.
We assume that the star-formation rate is simply proportional to the UV luminosity at 1600~\AA. Figure~\ref{madauplot} shows the star-formation rate density at z$>$2 inferred by \cite{bou09} uncorrected for dust extinction (black asterisks)
and corrected according to Eq.~\ref{eq_bowens} (green squares), i.e. by assuming the local relation between extinction and spectral slope.
The left hand panel is for luminous sources ($\rm L>0.3~L^*_{z=3}$), while the right hand panel is for less luminous sources ($\rm L>0.04~L^*_{z=3}$).
The red hatched area shows the star-formation rate corrected by using Eq.~\ref{eq_gallerani} (i.e. by assuming the MEC attenuation curve)
and the observed spectral slopes given by \cite{bou09}. The width of the hatched area reflects the uncertainties in the spectral slopes.
As expected, the star-formation rate density
after applying the correction obtained by using the MEC attenuation curve is substantially higher than
previous estimates based on local calibrations of the dust absorption correction. With the MEC-attenuation correction, the decline in the SFR density
is less pronounced than previous estimates.

We note that if the screen approximation holds for young high-z sources \citep[as suggested by ][]{reddy10}, which are possibly associated with dusty winds acting
as foreground screens, then the magnitude of the differential correction on the SFR density would be much larger than estimated above. In the
case of foreground screen-like absorption, one should indeed use the extinction curves (and not the attenuation curves), and the difference
between MEC and SMC should be much larger than for the corresponding attenuation curves.

We caution that we do not intend to provide a new measurement of the SFR at $z>4$, but simply to show how the actual SFR may be underestimated by adopting a relationship
between spectral slope and extinction inferred for local or intermediate redshift galaxies (Eq.~\ref{eq_bowens}).
As mentioned above, the careful derivation of the SFR by using a different prescription for the dust extinction requires a new, dedicated modeling of individual galaxies
at z$>$4,
possibly by including additional photometric (or even spectroscopic) measurements, which will probably become possible only with the next generation of facilities 
(JWST, TMT, EELT). Here, we note that our results may skew the peak of the SFR density from z~$\sim 2-3$ (Hopkins \& Beacom 2006) towards higher redshifts.

Our results may imply that the sources responsible for the reionization (or for keeping the universe reionized), and currently not detected in the high-$z$ surveys,
are not below the detection limit because they are intrinsically faint, but because they have been absorbed by dust. The knowledge of dust properties in the early Universe is therefore fundamental to studies of high-$z$ galaxies during reionization \citep[see also][]{dayal10a, dayal10b}.

\section{Conclusions}

We have investigated the extinction properties of dust at 3.9$<$z$<$6.4 by using
optical-near infrared spectra of 33 quasars in this redshift range. Our main goal has been to
investigate whether the SMC extinction curve, which accurately describes quasar reddening
at $z<2.2$, also represents a good prescription at these epochs.

We have fitted the quasar spectra with templates by varying the intrinsic spectral slope and by using
a large grid of extinction curves, including the well-known empirical laws and several theoretical
extinction curves (as well as mixtures of these extinction curves with the SMC).
We find that 7 quasars
in our sample require substantial ($0.8\geq A_{3000}\geq 2.0$) extinction. The best-fit extinction curve was found to consistently differ from the SMC, although the latter is still (marginally) consistent with the individual spectra. The best-fit extinction curve is generally flatter than the SMC in the UV, although there is a large spread among the different
extinction curves. We note that most of the dispersion among the best-fit extinction curves originate in the
mixture of BAL and non-BAL quasars. BAL quasars are always characterized by flatter extinction curves relative to non-BAL quasars.
If BAL reddened quasars are fitted simultaneously then the deviation from the SMC becomes highly significant ($> 2\sigma$).
The extinction curves characterizing non-BAL quasars have instead properties that are intermediate between BAL quasars and the SMC, and the
deviation from the latter is much less significant. We suggest that reddened BAL quasars at high-z are in an earlier evolutionary stage (in agreement with
previous studies) when the first generation of dust is produced;
then, as quasars evolve into non-BAL reddened systems their dust extinction curve properties approach the SMC extinction.
However, the higher confidence with which BAL quasars deviate from the SMC is probably also caused by BAL quasars in general being more reddened and therefore their extinction curve being able to be measured with higher accuracy.

We infer a mean extinction curve (MEC) for all reddened quasars at z$>$4, and by also dividing them into BAL and non-BAL.
According to the scenario discussed above, the MEC
of BAL quasars should represent the most appropriate description of the extinction curve in early Universe galaxies.

The finding that the extinction curves at z$>$4 deviate from those observed at lower redshift suggests that the dominant dust production mechanism
at these early epochs differs from that of local or intermediate redshift objects. At these early epochs, there is probably an higher contribution of dust by SNe than AGB stars. This is in line with the extinction curves theoretically
expected for SN-like dust providing generally the best fit to the quasar spectra at z$>$4. 

\cite{val09} demonstrated that the relative contribution of AGBs and SNe to the dust production in quasar is strongly dependent on the star-formation history of the host galaxy. In particular, the SNe contribution dominantes until the last major burst of star-formation. This scenario is supported by our finding of a more pronounced deviation from the SMC in the case of BAL quasars. In fact, since BAL quasars are probably younger than no-BAL quasars, the former may be closer to the starburst phase than the latter.   

We suggest that the extinction curve inferred for quasars at z$>$4 may also apply to young galaxies at similar redshifts. More specifically, we suggest that
the average change in the spectral slope for galaxies at z$>$4 (getting bluer) may reflect the same variation in extinction curve observed in quasars, rather than
a lower dust content. We infer the attenuation curve associated with the MEC of BAL quasars. The relationship between  
galaxy spectral slope $\beta$ and extinction $\rm A_{1600}$ used for local and intermediate redshift galaxies is modified using the MEC attenuation curve, which
we regard to be more appropriate at z$>$4. We use this relation to correct the star-formation rate density at z$\sim$4--6 for dust extinction, leading to substantially higher estimates than derived previously. 

\section*{Acknowledgment}
This work was partially supported by INAF and by ASI 
through contract ASI-INAF I/016/07/0. We are grateful to the anonymous referee for her/his useful comments. We thank D. Mortlock for providing us with the spectrum of ULAS J0203+0012 and H. Hirashita for the SNe-like theoretical extinction curves. We are thankful to A. Ferrara, R. Valiante and to all DAVID\footnote{http://wiki.arcetri.astro.it/bin/view/DAVID/WebHome} members for enlightening discussions.

\bibliographystyle{aa}

\appendix\label{app1}
\section{Results for unreddened quasars}
\begin{figure*}
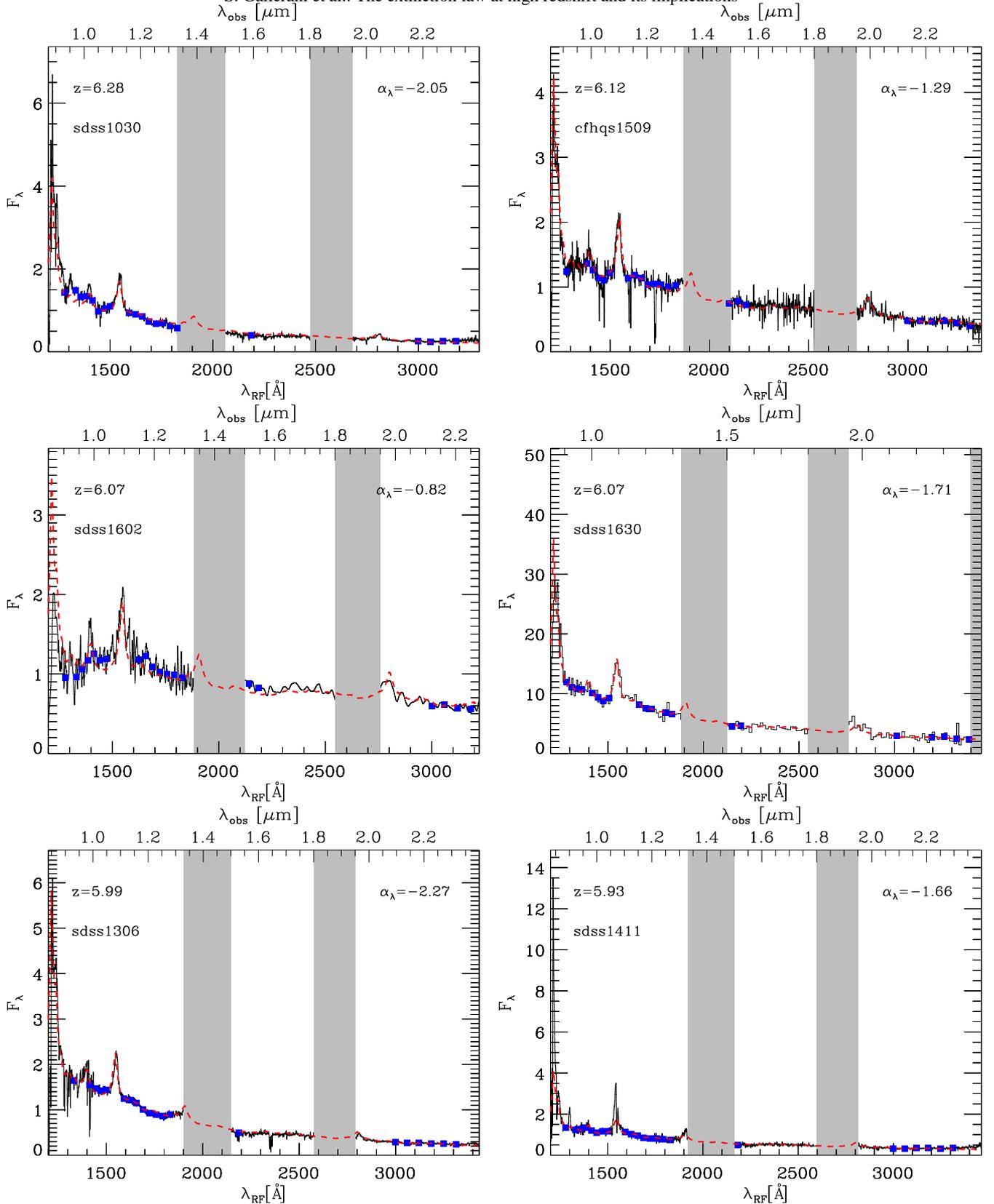

\centerline{
\psfig{figure=14721fg27.ps,width=9.5cm,angle=0}
$\!\!\!\!\!\!\!\!\!\!\!$
\psfig{figure=14721fg28.ps,width=9.5cm,angle=0}
}
\vspace{0.5cm}
\centerline{
\psfig{figure=14721fg29.ps,width=9.5cm,angle=0}
$\!\!\!\!\!\!\!\!\!\!\!$
\psfig{figure=14721fg30.ps,width=9.5cm,angle=0}
}
\vspace{0.5cm}
\centerline{
\psfig{figure=14721fg31.ps,width=9.5cm,angle=0}
$\!\!\!\!\!\!\!\!\!\!\!$
\psfig{figure=14721fg32.ps,width=9.5cm,angle=0}
}
\caption{Unreddened quasar spectra. The observed/rebinned spectra are indicated by black lines/filled blue squares, while the red dashed lines represent the best fitting spectra resulting from the $\chi^2$ analysis. The gray bands report regions of bad atmospheric transmission. In each panel, the name and the redshift of the quasar is shown at the top left, while the best fitting slope is reported at the top right.}
\label{unred1}
\end{figure*}

In Fig.\ref{unred1}--\ref{unred2}, we show the spectra of all the unreddened quasars (black solid line)
along with their best fitting template (red dashed line).
As for the reddened quasars, blue squares show the rebinned spectral value used for the fitting procedure.
In Table~\ref{tabapp1}, we report the fitting results
for quasars that do not require substantial reddening. We note that there are some
objects for which the ``no extinction'' case does not provide a good fit, although they are considered unreddened, since the
best-fit extinction curve is also characterized by a $P(\chi_{BF}^2>68\%)$. This may be due to both the actual extinction
curve not being part of our grid and some problems in the data. Since the grid of extinction curves is very extended, we plan to
reobserve these object to improve our analysis.


\begin{table}
\begin{center}
\caption{Unreddened quasars}
\label{tabapp1}
\begin{tabular}{l l l l}
\hline
 Name             &  z    &  $\alpha_{\lambda,Noext}$ & $P(\chi^2_{Noext})$\\
\hline
SDSS J1030 & 6.28 & -2.05 & 1.00\\
CFHQS 1509$^{\mathrm{a}}$ &  6.12&-1.29 & 0.71\\
SDSS J1602 &  6.07 & -0.82 & 1.00\\
SDSS J1630 &  6.06 & -1.71 &0.08  \\
SDSS J1306 &  5.99   & -2.27&0.19 \\
SDSS J1411 &  5.93   & -1.66& 0.88\\
SDSS J0836 &  5.80 & -1.87 & 0.97\\
SDSS J0005 &  5.85&-1.42 & 1.00\\
SDSS J0002 &  5.80 & -1.60 &0.06\\
SDSS J1044$^{\mathrm{a}}$ & 5.78 & -1.65&0.32\\
SDSS J02313&  5.41  & -2.71 & 0.33\\
SDSS J1614 &  5.31& -1.22 & 1.00\\
SDSSp J1208&  5.27  & -2.62&0.59\\
SDSS J0957&  5.16  &  -2.66& 0.36\\
SDSS J0756$^{\mathrm{b}}$ &   5.05&-1.64& 0.98\\
SDSS J1204&  5.05  &  -2.10& 0.24\\
SDSS J0017$^{\mathrm{a}}$ &  5.01  & -1.95 & 0.10\\
SDSS J0338&  5.00   &  -1.87& 0.42\\
SDSS J2216& 4.99    &  -1.49& 0.45\\
SDSS J1605$^{\mathrm{a}}$$^{\mathrm{, c}}$ &  4.92 &-1.33 & 1.00\\
SDSS J2200&  4.77   &  -1.37& 0.48\\
SDSS J0120$^{\mathrm{a}}$ &  4.73 & -1.21 & 1.00\\
SDSSp J1021&  4.70  &  -1.01&0.86\\
SDSS J1603 &  4.39 & -1.78 & 0.82\\
SDSS J0156$^{\mathrm{a}}$ &  4.32  & -1.22& 0.91\\
SDSS J0239$^{\mathrm{a}}$ &  4.02 & -1.39 & 1.00\\
\hline
\end{tabular}

$^{\mathrm{a}}$ BAL quasars;\\ $^{\mathrm{b}}$ For this spectrum we find a $P(\chi^2)\sim 0.6$ in the case of reddening by an extinction curve given by 30\% SMC + 70\% PISN2; \\$^{\mathrm{c}}$ For this spectrum we find a $P(\chi^2)\sim 0.6$ in the case of reddening by an extinction curve given by 70\% SMC + 30\% H2; 
\end{center}
\end{table}

\begin{figure*}
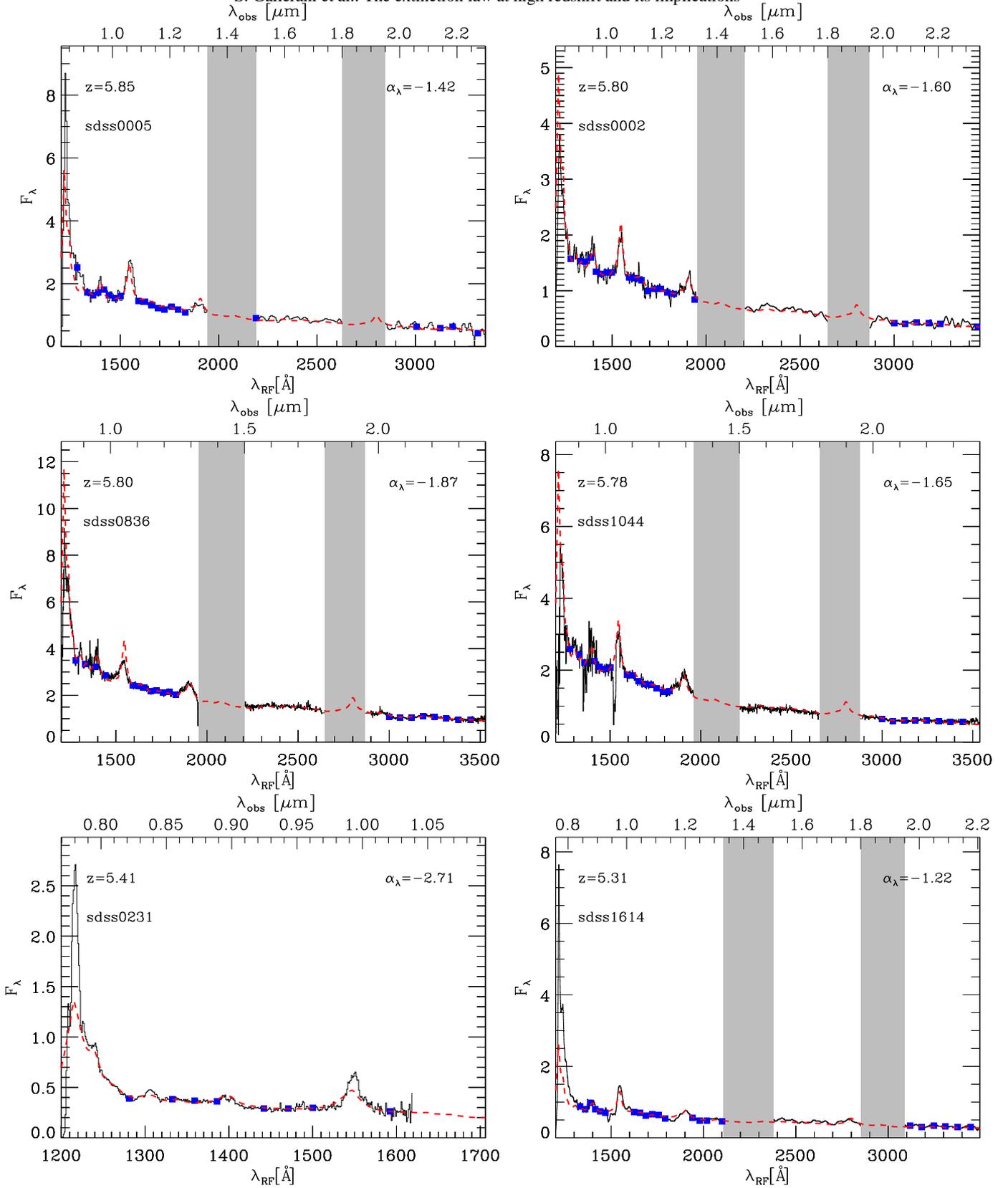

\centerline{
\psfig{figure=14721fg33.ps,width=9.5cm,angle=0}
$\!\!\!\!\!\!\!\!\!\!\!$
\psfig{figure=14721fg34.ps,width=9.5cm,angle=0}
}
\vspace{0.5cm}
\centerline{
\psfig{figure=14721fg35.ps,width=9.5cm,angle=0}
$\!\!\!\!\!\!\!\!\!\!\!$
\psfig{figure=14721fg36.ps,width=9.5cm,angle=0}
}
\vspace{0.5cm}
\centerline{
\psfig{figure=14721fg37.ps,width=9.5cm,angle=0}
$\!\!\!\!\!\!\!\!\!\!\!$
\psfig{figure=14721fg38.ps,width=9.5cm,angle=0}
}
\caption{Fig. A.1 continued.}
\end{figure*}

\begin{figure*}
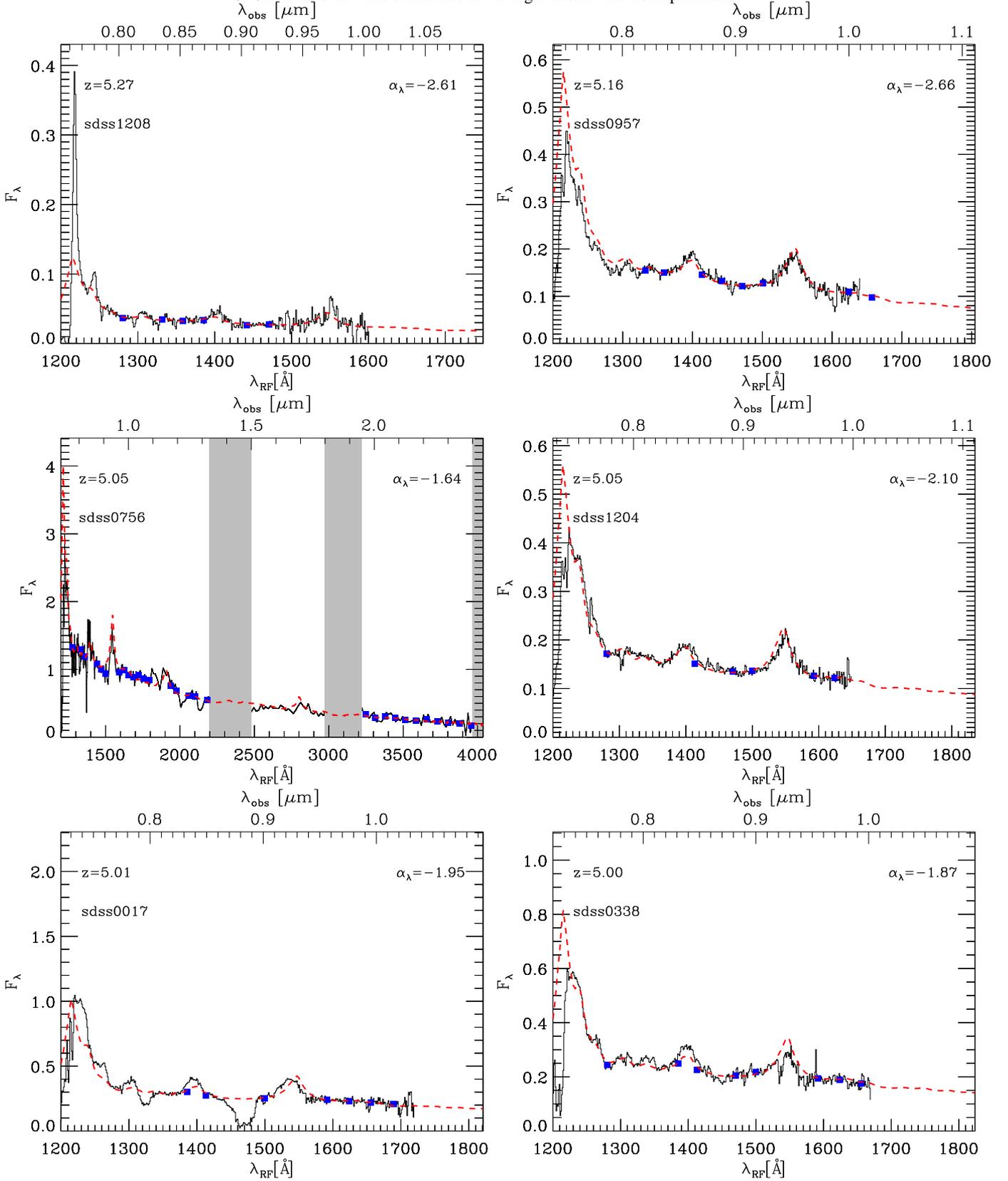

\centerline{
\psfig{figure=14721fg39.ps,width=9.5cm,angle=0}
$\!\!\!\!\!\!\!\!\!\!\!$
\psfig{figure=14721fg40.ps,width=9.5cm,angle=0}
}
\vspace{0.5cm}
\centerline{
\psfig{figure=14721fg41.ps,width=9.5cm,angle=0}
$\!\!\!\!\!\!\!\!\!\!\!$
\psfig{figure=14721fg42.ps,width=9.5cm,angle=0}
}
\vspace{0.5cm}
\centerline{
\psfig{figure=14721fg43.ps,width=9.5cm,angle=0}
$\!\!\!\!\!\!\!\!\!\!\!$
\psfig{figure=14721fg44.ps,width=9.5cm,angle=0}
}
\caption{Fig. A.1 continued.}
\end{figure*}

\begin{figure*}
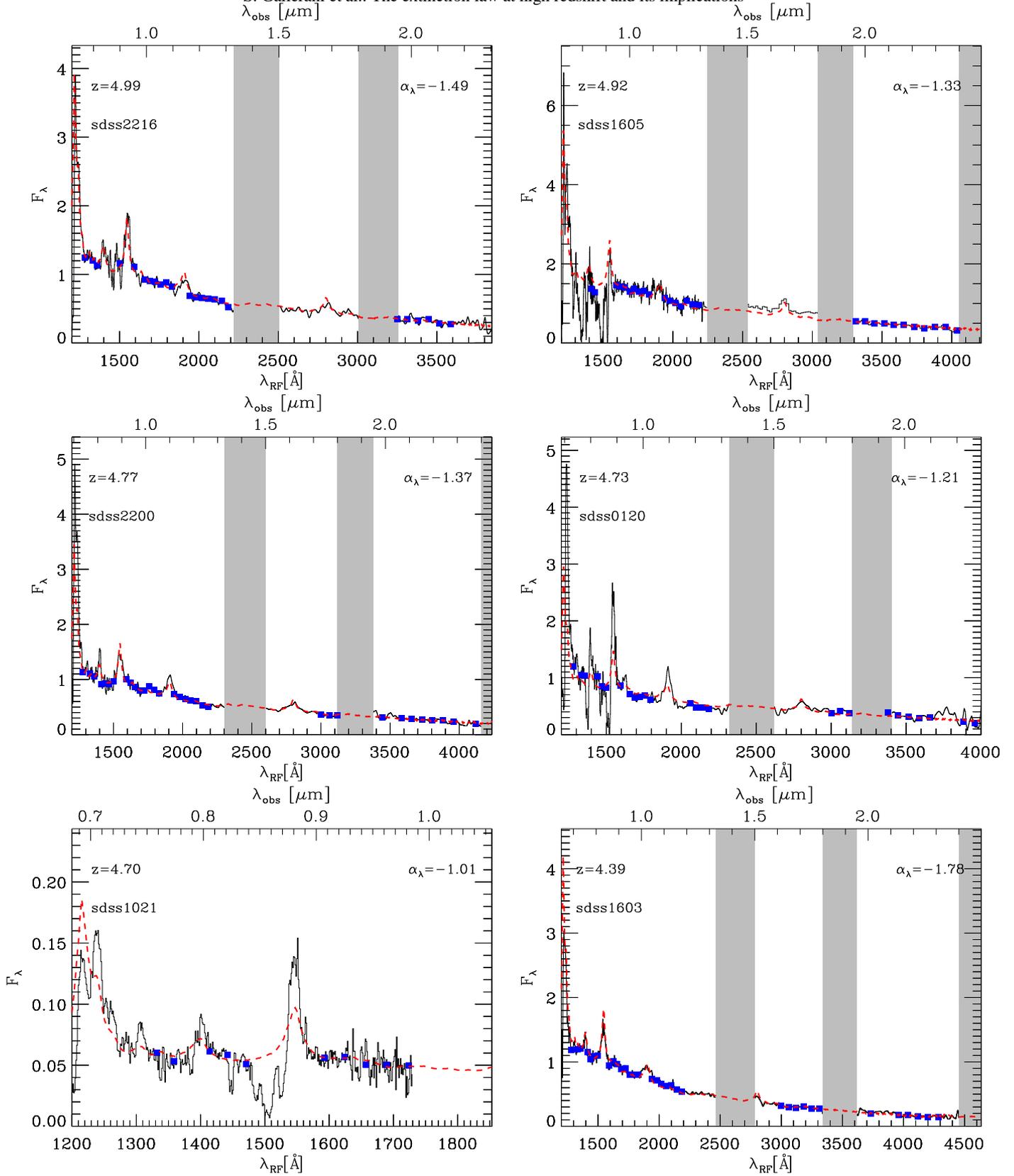

\centerline{
\psfig{figure=14721fg45.ps,width=9.5cm,angle=0}
$\!\!\!\!\!\!\!\!\!\!\!$
\psfig{figure=14721fg46.ps,width=9.5cm,angle=0}
}
\vspace{0.5cm}
\centerline{
\psfig{figure=14721fg47.ps,width=9.5cm,angle=0}
$\!\!\!\!\!\!\!\!\!\!\!$
\psfig{figure=14721fg48.ps,width=9.5cm,angle=0}
}
\vspace{0.5cm}
\centerline{
\psfig{figure=14721fg49.ps,width=9.5cm,angle=0}
$\!\!\!\!\!\!\!\!\!\!\!$
\psfig{figure=14721fg50.ps,width=9.5cm,angle=0}
}
\caption{Fig. A.1 continued.}
\label{unred2}
\end{figure*}
\begin{figure*}
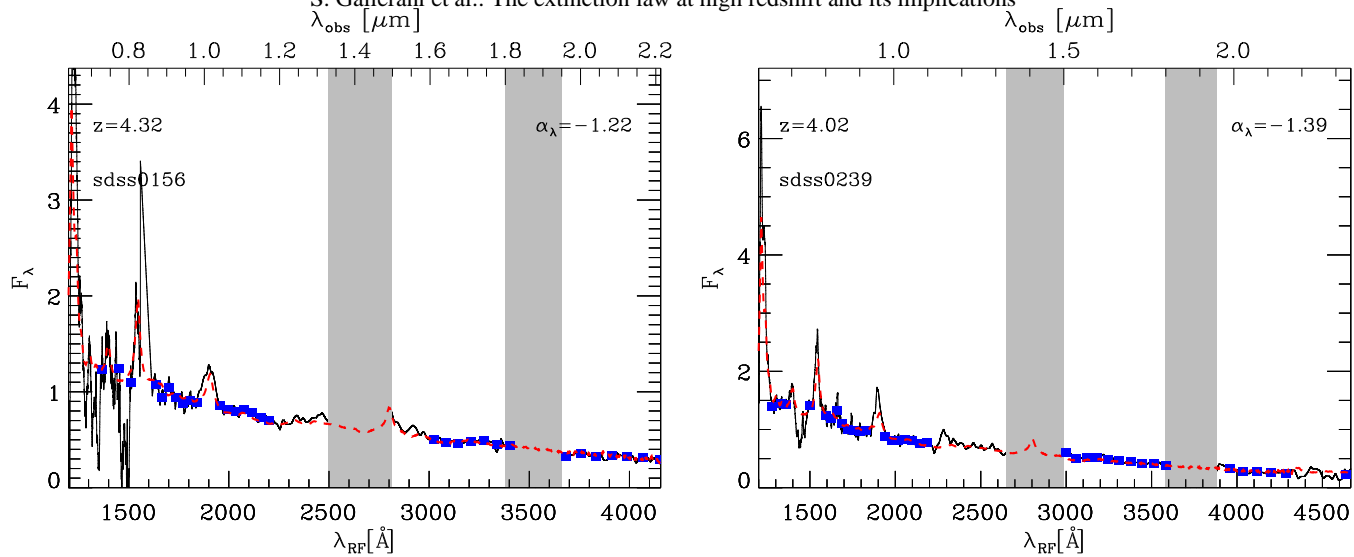

\centerline{
\psfig{figure=14721fg51.ps,width=9.5cm,angle=0}
$\!\!\!\!\!\!\!\!\!\!\!$
\psfig{figure=14721fg52.ps,width=9.5cm,angle=0}
}
\caption{Fig. A.1 continued.}
\label{unred2}
\end{figure*}
\end{document}